\documentclass[twocolumn]{autart}    

\makeatletter
\def\@journal{a journal}
\makeatother

\usepackage[longnamesfirst]{natbib}

\usepackage{graphicx}      
\usepackage{amsmath}
\usepackage{transparent}
\usepackage{cite}
\usepackage{bbold}
\usepackage{amssymb}
\usepackage{enumerate}
\usepackage{accents}
\usepackage{comment,color}
\usepackage{tikz}
 
 \usetikzlibrary{arrows}
 \usetikzlibrary{fit}
 \usetikzlibrary{calc}
  \pgfdeclarelayer{background1}
  \pgfdeclarelayer{background2}
  \pgfdeclarelayer{background3}
  \pgfdeclarelayer{background4}
  \pgfdeclarelayer{foreground}
  \pgfsetlayers{background4,background3,background2,background1,main,foreground}
  \tikzset{errorstyle/.style={thick,red,solid}}
  \tikzset{yrefstyle/.style={thick,black,dashed}}
  \tikzset{safetystyle/.style={thick,blue,dotted}}
  \tikzset{funnelstyle/.style={thick,blue,densely dotted}}
  \tikzset{funnelbackground/.style={black!20,opacity=0.5}}
  \tikzset{funneldstyle/.style={thin,blue,dashed}}
  \tikzset{qstyle/.style={green!50!black,ultra thick}}
  \tikzset{qhelpstyle/.style={green!50!black,very thin}}
  \tikzset{funnelfillstyle/.style={blue!20!white,opacity=0.8}}
  \tikzset{safetyfillstyle/.style={blue,opacity=0.1}}
  \tikzset{funnelthinfillstyle/.style={blue!5!white,opacity=0.8}}
  \tikzset{safetythinfillstyle/.style={blue!50!white,opacity=0.1}}
  
\newtheorem{thm1}{\bf Theorem}

\newtheorem{lem1}[thm1]{\bf Lemma}

\newtheorem{assum1}{\bf Assumption}
\newtheorem{rem1}{\bf Remark}
\newtheorem{exam1}{\bf Example}

\newenvironment{exam}{\begin{exam1}\rm}{\end{exam1}}

\newcommand*{\QEDB}{\hfill\ensuremath{\square}}
\newcommand*{\QEDD}{\hfill\ensuremath{\diamond}}

\newcommand{\R}{\mathbb{R}}
\newcommand{\N}{\mathbb{N}}

\newcommand{\cF}{\mathcal{F}}
\newcommand{\cI}{\mathcal{I}}
\newcommand{\cJ}{\mathcal{J}}

\newcommand{\cK}{\mathcal{K}}
\newcommand{\cL}{\mathcal{L}}
\newcommand{\cN}{\mathcal{N}}
\newcommand{\cM}{\mathcal{M}}

\newcommand{\eps}{\varepsilon}

\newcommand{\setdef}[2]{\left\{#1\,\left|\,\vphantom{#1} #2\,\right.\!\!\right\}}

\definecolor{antiquefuchsia}{rgb}{0.57, 0.36, 0.51}
\definecolor{auburn}{rgb}{0.43, 0.21, 0.1}

\usepackage[normalem]{ulem}

\allowdisplaybreaks[3]

\begin{document}

\begin{frontmatter}

\title{Synchronization with prescribed transient behavior: Heterogeneous multi-agent systems under funnel coupling\thanksref{footnoteinfo}}
\vspace{-4mm}
\subtitle{\large Extended arXiv version}

\thanks[footnoteinfo]{This work was supported by the National Research Foundation of Korea(NRF) grant funded by the Korea government(Ministry of Science and ICT) (No. NRF-2017R1E1A1A03070342).
This work was done while the first author was with Seoul National University.
}

\author[JGLee]{Jin Gyu Lee}\ead{jgl46@cam.ac.uk}, 
\author[STrenn]{Stephan Trenn}\ead{s.trenn@rug.nl},   
\author[SNU]{Hyungbo Shim}\ead{hshim@snu.ac.kr}  

\address[JGLee]{Control Group, Department of Engineering, University of Cambridge, United Kingdom} 
\address[STrenn]{Department of Mathematics, University of Groningen, Netherlands}  
\address[SNU]{ASRI, Department of Electrical and Computer Engineering, Seoul National University, Korea}        
          
\begin{keyword}     
synchronization, heterogeneous multi-agents, emergent dynamics, funnel control
\end{keyword}                     

\begin{abstract}
In this paper, we introduce a nonlinear time-varying coupling law, which can be designed in a fully decentralized manner and achieves approximate synchronization with arbitrary precision, under only mild assumptions on the individual vector fields and the underlying (undirected) graph structure.
The proposed coupling law is motivated by the so-called funnel control method studied in adaptive control under the observation that arbitrary precision synchronization can be achieved for heterogeneous multi-agent systems by a high-gain coupling; consequently we call our novel synchronization method `(node-wise) funnel coupling.'
By adjusting the conventional proof technique in the funnel control study, we are even able to obtain asymptotic synchronization with the same funnel coupling law.
Moreover, the emergent collective behavior that arises for a heterogeneous multi-agent system when enforcing arbitrary precision synchronization by the proposed funnel coupling law, is analyzed in this paper.
In particular, we introduce a single scalar dynamics called `emergent dynamics' which describes the emergent synchronized behavior of the multi-agent system under funnel coupling.
Characterization of the emergent dynamics is important because, for instance, one can design the emergent dynamics first such that the solution trajectory behaves as desired, and then, provide a design guideline to each agent so that the constructed vector fields yield the desired emergent dynamics.
We illustrate this idea via the example of a distributed median solver based on funnel coupling.
\end{abstract}

\end{frontmatter}

\section{Introduction}

\subsection{Synchronization of multi-agent systems}

During the last decade, synchronization and collective behavior of a multi-agent system have attracted increasing attention because of numerous applications in diverse areas, e.g., biology, physics, and engineering.
Initial studies focused on identical multi-agents \citep{olfati2004consensus,moreau2004stability,Ren05,seo2009consensus}, but the interest soon transferred to the heterogeneous case motivated by the fact that uncertainty, disturbance, and noise are prevalent in practice.
Earlier results in this direction such as \citep{wieland2013synchronous} have found that for synchronization to happen in a heterogeneous network, each agent must contain a common internal model.
However, recalling that heterogeneity may arise by noises or parameter perturbations, the assumption of common internal model may be too ideal, and approximate (practical) synchronization has been studied as an alternative \citep{montenbruck2015practical, ha2015practical}.
We want to note that only recently the emergence of collective behavior for heterogeneous multi-agent systems that achieve approximate synchronization is discussed, and  some attempts are made to analyze this behavior \citep{kim2016robustness,panteley2017synchronization,jglee18automatica}.

In this respect, a number of papers have considered the construction of a local controller to achieve arbitrary precision approximate synchronization (or asymptotic synchronization) for heterogeneous multi-agent systems.
In particular, output regulation theory, backstepping method, high-gain observer, adaptive control, and optimal control have been utilized.
Meanwhile, to the best of our knowledge, these works either have a common internal model assumption~\citep{de2012internal,isidori2014robust,modares2017optimal,casadei2017multipattern}, use sufficiently small (or large) parameters which depend on the global information such as the network topology~\citep{su2014cooperative,montenbruck2015practical,zhang2016almost,kim2016robustness,panteley2017synchronization}, need additional communication channels~\citep{lee2018practical,su2019semi}, or assume individual stability in the broad sense such as passivity~\citep{Arca07,delellis2015convergence}.

\subsection{Novel funnel coupling law and system class}

In this paper, we introduce a novel nonlinear time-varying coupling law, which overcomes the above mentioned restrictions, in particular, which
\begin{itemize}
\item can be designed in a fully decentralized manner, especially without the need of any global information such as the vector fields of other agents or the structure of communication graph,
\item does not require any additional assumptions on the individual vector fields such as stability in the broad sense or the common internal model assumption,
\item does not need additional communication and uses only the given diffusive coupling terms,
\item and achieves prescribed performance, in particular, uniform approximate synchronization with arbitrary precision.
\end{itemize}
For the set $\cN := \{1, \dots,  N\}$ of agents, the individual dynamics for agent $i\in\cN$ are assumed to be given by
\begin{subequations}\label{eq:eachdyn}
\begin{align}
\dot{x}_i(t) &= f_i(t, x_i(t)) + u_i(t, \nu_i(t)),  \label{eq:eachdyn_xidot}\\
\nu_i(t) &= \sum_{j \in \mathcal{N}_i} \alpha_{ij}\cdot (x_j(t) - x_i(t)), \label{eq:eachdyn_nui}
\end{align}
\end{subequations}
where $\mathcal{N}_i$ is a subset of $\mathcal{N}$ whose elements are the indices of the agents that send the information to agent~$i$.
The coefficient $\alpha_{ij}$ is the $ij$-th element of the adjacency matrix that represents the given graph.

\begin{assum1}\label{assum:graph}
{\rm (graph)} The communication graph induced by the adjacency element $\alpha_{ij}$ is undirected and connected, and thus, the Laplacian matrix $\mathcal{L}$ is symmetric, having one simple eigenvalue of zero. \QEDD
\end{assum1}

In the description, the (scalar) state at time $t \in \mathbb{R}$ is represented by $x_i(t) \in \mathbb{R}$, and $u_i : [t_0, \infty) \times \R \to \R$ is the nonlinear time-varying coupling law to be presented later on, which is a continuous mapping from the diffusive error term $\nu_i$ to the control input and is possibly time-varying.
Note that for the control design only the knowledge of the diffusive term \eqref{eq:eachdyn_nui} is assumed and that neither the knowledge of the agent's own state $x_i$ nor the neighbors' state $x_j$ is required.
It is a {\it heterogeneous} multi-agent system in the sense that each agent $i$ has its own vector field $f_i$.
We assume the following properties on the open loop dynamics of each agent.

\begin{assum1}\label{assum:fi}
{\rm (vector field)} For each $i\in\cN$, the function $f_i:[t_0, \infty) \times \mathbb{R} \to \mathbb{R}$ is measurable in $t$, locally Lipschitz with respect to $x_i$, and bounded on each compact subset of $\R$ uniformly in $t \in [t_0, \infty)$. \QEDD
\end{assum1}

Note that the time-varying $f_i$ can include an external input, a disturbance, and/or noise as well, and we do not assume any stability of the node dynamics $\dot x_i = f_i(t,x_i)$.

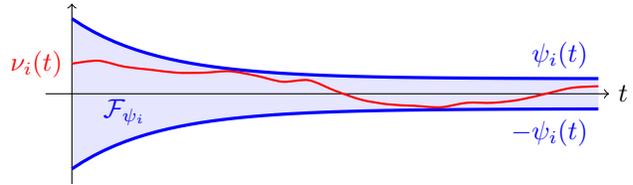
\begin{figure}[hbt]
\centering
\begin{tikzpicture}[baseline = 0,xscale=0.7,yscale=0.4,variable=\t,domain=0:10,samples=50]
   \fill[fill = blue!10] plot ({10-\t},{2/exp(0.6*(10-\t))+0.5}) -- plot (\t,{-2/exp(0.6*\t)-0.5});
   \draw[draw=black,->]  (0,-3) -- (0,3);
   \draw[draw=black,->]  (-0.5,0) -- (10.2,0) node[anchor = west] {$t$};
   \draw (12,0) node {};
   \draw[color=blue,very thick] plot (\t,{2/exp(0.6*\t)+0.5}) node[anchor = south east] {$\psi_i(t)$};
   \draw[color=blue,very thick] plot (\t,{-2/exp(0.6*\t)-0.5}) node[anchor = north east] {$-\psi_i(t)$};
   \draw[errorstyle] plot[smooth] coordinates {(0,1) (0.5,1.1) (1,0.9) (1.5,0.8) (2,0.7) (2.5,0.7) (3,0.75) (3.5,0.6) (4,0.4) (4.5,0.45) (5,0.1) (5.5,-0.2) (6,-0.35) (6.5,-0.4) (7,-0.45) (7.5,-0.3) (8,-0.3) (8.5,-0.2) (9,0) (9.5,0.2) (10,0.25)};
   \draw[blue] (1,-0.6) node {$\cF_{\psi_i}$};
   \draw (0,1)  node[red,left] {$\nu_i(t)$};
\end{tikzpicture}
\caption{Prescribed performance: the diffusive error $\nu_i$ evolves within the funnel $\cF_{\psi_i}$.}\label{fig:funnel}
\end{figure}

Our coupling law to ensure synchronization with prescribed performance is inspired by the so-called funnel controller \citep{ilchmann2002tracking}.
Given the desired time-varying function $\psi_i$ for each agent~$i$, our goal is to ensure that the diffusive error term~$\nu_i$ of \eqref{eq:eachdyn_nui} evolves within the funnel
\[
   \cF_{\psi_i} := \setdef{(t,\nu_i)}{|\nu_i|<\psi_i(t)},
\]
as in Figure~\ref{fig:funnel}.

In fact, we can achieve uniform, arbitrary precision approximate synchronization in the sense that for any arbitrarily small $\eta>0$ and any given bounded set of initial values, we can easily choose funnel boundaries $\psi_i$ such that $|\nu_i(t_0)|<\psi_i(t_0)$ and $\limsup_{t \to \infty} \psi_i(t) \leq \eta$ which will, using our proposed method, result in $\limsup_{t \to \infty}|\nu_i(t)| \le \eta$ for all $i \in \cN$ with a uniform convergence rate (given by the shape of the funnels).
Note that the property $\limsup_{t \to \infty}|\nu_i(t)| \le \eta$ implies $\limsup_{t \to \infty} \|\cL x(t)\|_\infty \le \eta$ which in turn, by Assumption \ref{assum:graph}, implies that
\begin{equation}\label{eq:x_i-x_j}
   \limsup_{t\to\infty} |x_i(t)-x_j(t)| \le \frac{2\sqrt{N}}{\lambda_2} \eta, \quad \forall i,j\in\cN, 
\end{equation}
where $\lambda_2$ is the algebraic connectivity of the graph (see \eqref{eq:a6}).

To achieve this control objective, we propose for each $i\in\cN$ the following \emph{(node-wise) funnel coupling} law
\begin{align}\label{eq:funnel_coup}
u_i(t, \nu_i) = \mu_i\left(\frac{\nu_i}{\psi_{i}(t)}\right) := \gamma_i\left(\frac{|\nu_i|}{\psi_{i}(t)}\right)\frac{\nu_i}{\psi_{i}(t)} \in \mathbb{R}
\end{align}
where the functions $\psi_i$ and $\gamma_i$ satisfy the following assumption.

\begin{assum1}\label{assum:funnel}
{\rm (funnel)} Each function $\psi_i : [t_0, \infty) \to \R_{ >0}$ is bounded and differentiable with bounded derivative; i.e., there are $\overline{\psi}>0$ and $\theta_{\psi}>0$ such that 
$$0<\psi_i(t)\leq\overline{\psi} \;\; \text{and} \;\; |\dot{\psi}_i(t)|\leq\theta_\psi, \quad \forall t\in[t_0,\infty), i \in \cN.$$ 
The gain functions $\gamma_i:[0,1)\to\R_{> 0}$, $i\in\cN$, are non-decreasing and satisfy $\lim_{s \to 1} \gamma_i(s) = \infty$. \QEDD
\end{assum1}

A possible (agent independent) choice for $\gamma_i$ and $\psi_i$ is
\[
   \gamma_i(s) = \frac{1}{1-s} 
   \text{ and } \psi_i(t) = (\overline{\psi}-\eta) e^{-(\theta_\psi/\overline{\psi})
   (t-t_0)} + \eta,
\]
where $\overline{\psi},\theta_\psi
> 0$ and $\eta\geq 0$.

This funnel coupling law is motivated by the observation that approximate synchronization with arbitrary precision can be obtained by the high-gain linear coupling law $u_i(t, \nu_i) = k\nu_i$ \citep{jglee18automatica} (which corresponds to the high-gain property in the funnel control study).

\subsection{Related approaches}

The idea of funnel coupling has been first proposed in \citep{shim2015preliminary}, however, due to some technical reasons, the analysis was conducted only for the weakly centralized funnel coupling, i.e., $u_i(t) = \max_j\gamma_j(|\nu_j|/\psi(t))\nu_i/\psi(t)$, and only when the underlying graph is $d$-regular with $d > N/2 - 1$, where $d$ is the degree of every node.
These technical limitations are resolved in this paper, and we can now consider fully decentralized coupling law \eqref{eq:funnel_coup} with an arbitrarily given graph which is undirected and connected.
This new approach also allows the performance functions $\psi_{i}$ to converge asymptotically to zero, i.e., $\lim_{t \to \infty} \psi_{i}(t) = 0$, by which we obtain {\em asymptotic} synchronization for heterogeneous multi-agent systems.
In particular, we have $\lim_{t\to \infty} \nu_i(t) = 0$, $i \in \cN$ by the fact that $|\nu_i(t)| < \psi_{i}(t)$ for all $t \ge t_0$ and $i \in \cN$.
This, in fact, seems to violate the common presumption, in the synchronization community, that heterogeneous multi-agent systems can not asymptotically synchronize without a common internal model.
This violation is resolved by observing that we use a \emph{time-varying} coupling law, which is not considered in the framework of the internal model principle for multi-agent systems \citep{wieland2013synchronous}.
In fact, unlike the internal model principle results, it is observed in this paper, that, as the performance function approaches zero, the coupling term approaches a possibly non-zero time-varying signal, which compensates the heterogeneity of the individual agents.
Specific use of this idea to solve distributed consensus optimization can be found in \citep{leeutility}.
We want to emphasize that even when asymptotic synchronization is achieved, the input $u_i(t, \nu_i(t))$ can still be bounded.
In fact, even though the performance functions $\psi_i$ are asymptotically converging to zero, the diffusive term $\nu_i$, which also converges asymptotically to zero, makes the fraction $\nu_i(t)/\psi_i(t)$ be strictly contained inside the interval $(-1, 1)$ uniformly, making the input $\mu_i(\nu_i(t)/\psi_i(t))$ to be uniformly bounded.
We refer to Section~\ref{sec:MR} for sufficient conditions that guarantee the boundedness of input.

Relying also on the observation that arbitrary precision synchronization can be achieved by the high-gain linear coupling law, a dynamic coupling law motivated by the $\lambda$-tracking studied in adaptive controls \citep{ilchmann1994universal} given, for instance, as
\begin{align*}
u_i(t, \nu_i(t)) &= k_i(t)\nu_i(t), \\
\dot{k}_i(t) &= \begin{cases} |\nu_i(t)|(|\nu_i(t)| - \eta_i) &\mbox{ if } |\nu_i(t)| > \eta_i, \\ 0 &\mbox{ otherwise},\end{cases}
\end{align*}
has been introduced in \citep{shafi2014adaptive, li2012consensus, lv2017novel, kim2017adaptation, lee2018practical}.
But, most of them considered homogeneous networks, and for a heterogeneous network, additional communication between the coupling gains $k_i$ has been introduced to ensure that the collective behavior of the network is as desired.
In fact, funnel control has advantages compared to $\lambda$-tracker
such as that the transient behavior can be directly controlled and that the gain is not monotonically increasing, and thus, does not amplify the measurement noise unnecessarily.

\subsection{Emergent dynamics}
To estimate the behavior of the network when synchronization is achieved in this way, as in \citep{kim2016robustness, panteley2017synchronization, jglee18automatica}, the emergent collective behavior that arises from the closed loop system \eqref{eq:eachdyn} with \eqref{eq:funnel_coup} is analyzed in this paper.
In particular, we introduce a single scalar dynamics which we call `emergent dynamics' (which depends on the individual vector fields $f_i$ and the functions $\mu_i$, $\psi_i$ for all $i \in \mathcal{N}$) that is capable of illustrating the emergent synchronized behavior of the whole network by its solution trajectory.

Characterization of the emergent dynamics is important, for instance, when synthesizing a heterogeneous network for some specific purposes.
In particular, one can design the emergent dynamics with the desired behavior, and then, provide a guideline to each agent (which allows fully decentralized design) so that the constructed $f_i$ and $\mu_i$ yield the desired emergent dynamics.
This scheme of constructing a heterogeneous network with the desired collective behavior is first introduced in \citep{jglee18automatica} and has many interesting applications, e.g., distributed state estimation, estimation of the number of agents, and economic dispatch problem.
For instance, it is analyzed that the emergent behavior of a heterogeneous network~\eqref{eq:eachdyn} under the high-gain coupling $u_i(t, \nu_i) = k\nu_i$, follows the `blended dynamics' given by 
\begin{align}\label{eq:blend}
\dot \xi = \frac{1}{N} \sum_{i=1}^N f_i(t, \xi).
\end{align}
Under this observation, in~\citep{jglee18automatica}, for example, a network that estimates the number of agents is designed as $\dot{x}_1 = -x_1 + 1 + k\nu_1$ and $\dot{x}_i = 1 + k\nu_i$ for $i \neq 1$, which has $\dot \xi = -(1/N)\xi + 1$ as its blended dynamics, i.e., the emergent collective behavior asymptotically converges to the number of agents~$N$.
Now, the emergent dynamics to be introduced later on takes clearly a different form compared to the blended dynamics, and by this difference, a new application might occur, which needs further inspection.
A particular example illustrating the utility of the emergent dynamics is given in Section \ref{subsec:app_dms} as a distributed median solver.

We emphasize that each agent may be unstable (or malfunctioning, or even malicious), as long as their combination, i.e., the emergent dynamics is stable, hence the stability of the group is maintained. In particular, in a large group of agents malicious agents cannot destabilize the overall system provided the majority of agents behaves ``good''.
Synchronization achieved in this way is also robust against external disturbance, noise, and/or uncertainty in the agent dynamics due to the stability of the emergent dynamics.

\subsection{Paper organization and notation}
The paper is organized as follows.
In Section~\ref{sec:MR}, it is proven that the proposed node-wise funnel coupling law achieves synchronization with respect to the given performance function.
Some sufficient conditions that ensure boundedness of the inputs are also given at the end of that section.
Section~\ref{sec:Eb} analyzes the emergent collective behavior that arises when enforcing synchronization by the proposed funnel coupling law.
Then, in Section~\ref{sec:disc}, we discover the properties of the emergent dynamics, and also discuss the possible application related to these properties.

{\em Notation:}
Laplacian matrix $\mathcal{L} = [l_{ij}] \in \mathbb{R}^{N \times N}$ of a graph is defined as $\mathcal{L} := \mathcal{D} - \mathcal{A}$, where $\mathcal{A} = [\alpha_{ij}]$ is the adjacency matrix of the graph and $\mathcal{D}$ is the diagonal matrix whose diagonal entries are determined such that each row sum of $\mathcal{L}$ is zero.
By its construction, it contains at least one eigenvalue of zero, whose corresponding eigenvector is $1_N := [1,\dots,1]^\top \in \R^N$, and all the other eigenvalues have non-negative real parts.
For undirected graphs, the zero eigenvalue is simple if and only if the corresponding graph is connected.
For vectors or matrices $a$ and $b$, ${\rm col}(a,b) := [a^\top,b^\top]^\top$.
For matrices $A_1, \dots, A_k$, we denote by $\text{diag}(A_1, \dots, A_k)$ the corresponding block diagonal matrix.
For a non-empty set $\Xi \subseteq \R$, $|x|_{\Xi}$ denotes the distance between the value $x \in \R$ and $\Xi$, i.e., $| x |_{\Xi} := \inf_{y \in \Xi} | x - y|$. 

\section{Heterogeneous multi-agent systems under node-wise funnel coupling}\label{sec:MR}

The intuition of the funnel coupling law \eqref{eq:funnel_coup} is simple, following that of funnel control, which is to increase the gain infinitely large as the diffusive error approaches the funnel boundary.
Then, the high-gain precludes boundary contact.
For instance, if agent $i$ has only one neighbor denoted as agent $j$, and if the difference between two agents, $\nu_i(t) = \alpha_{ij}(x_j(t) - x_i(t))$, approaches the funnel boundary $\pm \psi_i(t)$ so that $\psi_i(t) - |\nu_i(t)|$ becomes closer to zero, then the gain $\gamma_i(|\nu_i(t)|/\psi_{i}(t))$ gets larger towards infinity, and the state~$x_i$ will tend to its neighbor $x_j$ since the large coupling term dominates the vector field $f_i(t, x_i)$, and the error $\nu_i(t)$ will remain inside the funnel.
However, with more than one neighbor, this intuition becomes no longer straightforward because two neighbors of agent $i$ may attract $x_i$ in the opposite direction with almost infinite power.
In the following, we will prove that all the errors $\nu_i(t)$ remain inside the funnel, which is however far more complicated and also requires the following technical assumption, which guarantees that, as long as the diffusive error is contained in the funnel, i.e., $|\nu_i(t)| < \psi_i(t)$, finite time escape cannot occur.

\begin{assum1}\label{assum:finite}
{\rm (no finite time escape)}
The dynamical systems defined by
\begin{equation}\label{eq:uplowdyn}
\dot{\overline{\chi}}(t) = \max_{i \in \cN} f_i(t, \overline{\chi}(t)),\quad \dot{\underline{\chi}}(t) = \min_{i \in \cN} f_i(t, \underline{\chi}(t))
\end{equation}
have complete solutions $\overline{\chi}, \underline{\chi} : [t_0, \infty) \to \R$ for any initial values $\overline{\chi}(t_0), \underline{\chi}(t_0) \in \mathbb{R}$ and for any initial time $t_0$. \QEDD
\end{assum1}

We stress that if the functions $f_i$ are globally Lipschitz in $x_i$, then Assumption~\ref{assum:finite} holds.

\begin{lem1}\label{lem:finite} 
Under Assumptions \ref{assum:fi}, \ref{assum:funnel}, and \ref{assum:finite}, assume that a solution of the system \eqref{eq:eachdyn} with \eqref{eq:funnel_coup} exists on $[t_0, \omega)$ for a finite $\omega > t_0$ and satisfies $|\nu_i(t)| < \psi_i(t)$, for all $t \in [t_0, \omega)$ and $i \in \cN$.
Then, there exists $M > 0$, depending on $(t_0, \omega, x_1(t_0),\ldots x_N(t_0))$, such that $|x_i(t)| < M$, for all $t \in [t_0, \omega)$ and $i \in \cN$. \QEDD
\end{lem1}

Before providing the proof of Lemma~\ref{lem:finite}, we stress that the bound $M$ does not depend on the particular choice of $\psi_i$, and that the boundedness of $x_i(t)$ on a finite time interval is established without relying on the boundedness of $u_i(t,\nu_i(t))$ which can be unbounded when $\nu_i(t)$ approaches the funnel boundary $\psi_i(t)$.
These properties will be used in our main result of Theorem \ref{thm:funnel}.

\begin{pf}
For a solution $x:[t_0,\omega)\to\R^N$ of~\eqref{eq:eachdyn} with~\eqref{eq:funnel_coup}, choose a time-varying index $J(t) \in \cN$ such that $x_{J(t)}(t) = \max_i x_i(t)$ and  $\dot{x}_{J(t)}(t) \ge \dot{x}_k(t)$ for all those $k \in \cN$ with $x_k(t) = \max_i x_i(t)$.
Then, the upper right Dini derivative of $x_{J(t)}(t)$ denoted as $D^+x_{J(t)}(t)$ satisfies
\begin{align}\label{eq:dplus}
\begin{split}
&D^+x_{J(t)}(t)\le \dot{x}_{J(t)}(t) \\
&\quad = f_{J(t)}(t, x_{J(t)}(t)) + \gamma_{J(t)}\left(\frac{|\nu_{J(t)}(t)|}{\psi_{J(t)}(t)}\right)\frac{\nu_{J(t)}(t)}{\psi_{J(t)}(t)} \\
&\quad \le f_{J(t)}(t, x_{J(t)}(t)) \le \max_i f_i(t, x_{J(t)}(t))
\end{split}
\end{align}
where the second inequality follows from the fact that $\gamma_{J(t)}$ and $\psi_{J(t)}$ are non-negative and $\nu_{J(t)}(t)$ is non-positive, because $x_{J(t)}(t)$ is a maximum.
Hence, by Assumption~\ref{assum:finite}, there exists $M_+ > 0$, depending on $(t_0, \omega, x_1(t_0),\ldots x_N(t_0))$, such that $x_{J(t)}(t)$ is upper bounded by $M_+$ for $t \in [t_0, \omega)$.
Similarly, we can find $M_- > 0$ such that $\min_i x_i(t) \ge -M_-$ for all $t \in [t_0, \omega)$, which concludes the claim. \QEDB
\end{pf}

\begin{thm1}\label{thm:funnel}
Consider the system \eqref{eq:eachdyn} coupled via node-wise funnel coupling \eqref{eq:funnel_coup}.
Under Assumptions \ref{assum:graph}--\ref{assum:finite}, 
if the initial values $x_i(t_0)$ are such that $|\nu_i(t_0)| < \psi_i(t_0)$, for all $i \in \cN$, then the solutions $x_i(t)$ exist for all $t \ge t_0$ and satisfy
\begin{equation}\label{eq:goal}
|\nu_i(t)| < \psi_i(t), \quad \forall t \ge t_0, \; i \in \cN.
\end{equation}
From this, the inequality \eqref{eq:x_i-x_j} holds, and thus, approximate (when $\eta>0$ is small) or asymptotic (when $\eta=0$) synchronization is achieved. \QEDD
\end{thm1}

While the following proof will ensure \eqref{eq:goal}, we want to derive \eqref{eq:x_i-x_j} from \eqref{eq:goal} here because this will introduce an important matrix $R$ that will be used frequently.
Define a matrix $R \in \R^{N \times (N-1)}$ such that the matrix $[(1/\sqrt{N})1_N, R]$ becomes orthogonal and $[(1/\sqrt{N})1_N, R]^\top \mathcal{L} [(1/\sqrt{N})1_N, R] = {\rm diag}(0,\lambda_2,\dots,\lambda_N)$ with $0 < \lambda_2 \le \cdots \le \lambda_N$.
Then, we have the following properties:
\begin{itemize}
\item $\Lambda = R^\top \mathcal{L} R$ and $\mathcal{L} = R \Lambda R^\top$ where $\Lambda = {\rm diag}(\lambda_2,\dots,\lambda_N)$ 
\item $RR^\top = I_N - (1/N)1_N1_N^\top$, and thus, $R_i R^\top \boldsymbol{x} = x_i - x_s$ where $R_i$ is the $i$-th row of $R$, $\boldsymbol{x} = {\rm col}(x_1,\dots,x_N)$, and $x_s := (1/N)\sum_{i=1}^N x_i$.
\end{itemize}
Let $\mathcal{L}_\psi = {\rm diag}(1/\psi_1(t),\dots,1/\psi_N(t)) \mathcal{L}$.
Then,
\begin{align}\label{eq:a6}
\begin{split}
&\max_{i \in \mathcal{N}} |x_i- x_s| = \|RR^\top\boldsymbol{x}\|_\infty \\
&= \|R\Lambda^{-1}R^\top\text{diag}(\psi_1(t), \dots, \psi_N(t))\mathcal{L}_\psi(t)\boldsymbol{x}\|_\infty \\
&\le \sqrt{N}\|R\Lambda^{-1}R^\top\|_2 \left(\max_{i \in \mathcal{N}} \psi_i(t)\right) \|\mathcal{L}_\psi(t)\boldsymbol{x}\|_\infty \\
& \le \sqrt{N} \max_{i \in \mathcal{N}} \psi_i(t)/\lambda_2 
\end{split}
\end{align}
if the solution remains inside the funnel; that is, $1 > |\nu_i(t)|/\psi_i(t) = |\mathcal{L}_i\boldsymbol{x}(t)|/\psi_i(t) = |\mathcal{L}_{\psi,i}(t)\boldsymbol{x}(t)|$ in which the subscript $i$ means the $i$-th row.
Since $x_i-x_j = (x_i-x_s)-(x_j-x_s)$, \eqref{eq:x_i-x_j} follows.

\begin{pf}
The proof is done by a contradiction.
Suppose that there is a particular solution of \eqref{eq:eachdyn} and \eqref{eq:funnel_coup} such that the inequality in \eqref{eq:goal} holds only for a finite time interval $[t_0,\omega)$ and is violated at $t = \omega$.
This implies that there is a time sequence $\{\tau_k\}$ such that $\tau_k$ is strictly increasing and $\lim_{k\to\infty}\tau_k = \omega$, and 
\begin{align*}
\cI_+(\{\tau_k\}) &:= \left\{i \in \cN : \lim_{k \to \infty} \frac{\nu_i(\tau_k)}{\psi_{i}(\tau_k)} = 1 \right\} \text{is non-empty,} \\
\text{or} & \\
\cI_-(\{\tau_k\}) &:= \left\{i \in \cN : \lim_{k \to \infty} \frac{\nu_i(\tau_k)}{\psi_{i}(\tau_k)} = -1 \right\} \text{is non-empty.}
\end{align*}
Let us first assume that $\cI_+(\{\tau_k\})$ is non-empty.
We will first show that a contradiction occurs if $\cI_+(\{\tau_k\}) = \cN$.
If $\cI_+(\{\tau_k\}) \subsetneq \cN$, we will then show that it is possible to construct another time sequence $\{\bar\tau_k\}$ (based on  $\{\tau_k\}$), such that 
\begin{equation}\label{eq:inc}
|\cI_+(\{\tau_k\})| < |\cI_+(\{\bar\tau_k\})|
\end{equation}
where the notation $|\cdot|$ implies the cardinality of the set.
By repeating this argument (i.e., by replacing the role of $\{\tau_k\}$ with $\{\bar\tau_k\}$), we arrive after finitely many steps at the equality $\cI_+(\{\tau_k\}) = \cN$, which yields a contradiction.
This means that there is no such sequence $\{\tau_k\}$ that makes $\cI_+(\{\tau_k\})$ non-empty.
Similarly, it can be seen that there is no sequence that makes $\cI_-(\{\tau_k\})$ non-empty.
Therefore, we conclude there is no such finite time $\omega$ and the control objective \eqref{eq:goal} is achieved for all $t\geq t_0$.

Let us carry out the above described proof steps.
For convenience, we write $\cI$ instead of $\cI_+(\{\tau_k\})$ in the following. 
Let
\[
    W(t) := \sum_{i \in \cI} \nu_i(t) = \sum_{i \in \cI} \sum_{j \in \mathcal{N}} \alpha_{ij}\cdot (x_j(t) - x_i(t)).
\]
Note that, by the definition of $\cI$, for each $i \in \cI$, there exists a sufficiently large $k_i^* \in \mathbb{N}$ such that $\nu_i(\tau_k) > 0$ for all $k \ge k_i^*$ because $\psi_i(t)>0$ for all $t \in [t_0, \omega)$.
Hence there is $k$ such that $W(\tau_{k}) > 0$.
However, this inequality is violated if $\cI = \cN$ because
\[
W(t) = \sum_{i\in\cN}\sum_{j\in\cN} \alpha_{ij}(x_j(t)-x_i(t)) \equiv 0, \quad \forall t \in [t_0, \omega),
\]
which is a simple consequence from the general property of an undirected graph that for any index set $\cK\subseteq \cN$ and any vector $\chi = [\chi_i] \in \R^N$, 
\begin{equation}\label{eq:sum_sum=0}
    \sum_{i \in \cK} \sum_{j \in \cK} \alpha_{ij} (\chi_j - \chi_i) = 0.
\end{equation}

Hence we have shown that $\cI=\cN$ is not possible and we continue the proof for the case that $\cI \subsetneq \cN$.
For this purpose, note that $W(t)$ is continuously differentiable, $W(t) < \sum_{i \in \cI} \psi_i(t)$ on $[t_0,\omega)$, and $\lim_{k \to \infty} W(\tau_k) = \sum_{i \in \cI} \psi_i(\omega)$.
Let us now consider a strictly decreasing sequence $\{\eps_q\} \subseteq (0, 1)$ such that $\lim_{q \to \infty} \eps_q = 0$ and that $W(t_0) < (1- \eps_0)\sum_{i\in \cI}\psi_i(t_0) $. 
Choose a subsequence $\{\tau_{k_q}\}_{q\in\N}$ of $\{\tau_k\}$ such that
\begin{equation}\label{eq:jg1}
W(\tau_{k_q}) \ge \left( 1 - \frac{\eps_q}{2} \right) \sum_{i \in \cI} \psi_i(\tau_{k_q}),
\quad \forall q \in \N.
\end{equation}
Based on this subsequence, we now construct a sequence $\{s_q\}_{q\in\N}$ such that (see Figure~\ref{fig:s^n+1_p})
\begin{equation}\label{eq:jg2}
s_q\! :=\! \max\setdef{s\! \in\! [t_0, \tau_{k_q}]}{ W(s)\! =\! (1 - \eps_q) \sum_{i \in \cI}\psi_i(s)}\!.
\end{equation}

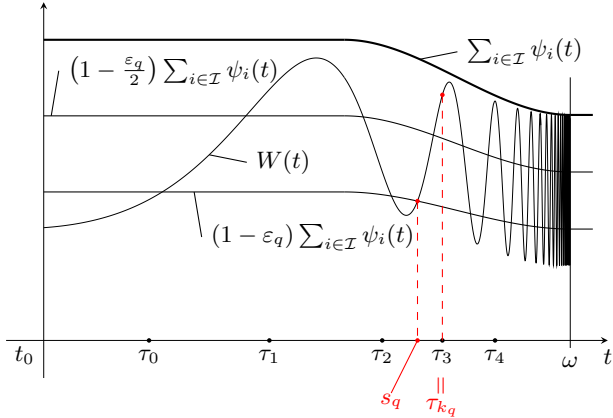
\begin{figure}[hbt]
\begin{tikzpicture}[>=stealth]
\draw[very thin] (0,-0.2) node[left] {\footnotesize $t_0$};
\draw[very thin,->] (0,-0.5) -- (0,4.5);
\draw[very thin,->] (-0.5,0) -- (7.5,0) node[below] {\footnotesize $t$};
\draw[very thin] (7,3.5) -- (7,-0.1) node[below] {\footnotesize $\omega$};
\draw[thick] (0,4) node[left] {\footnotesize} -- (4,4);
\draw[line width = 0.8, variable = \t, domain= 4 : 7, samples = 180]
plot[smooth] (\t, {2*\t^3/27 - 11*\t^2/9 + 56*\t/9 - 164/27});
\draw (5,3.7407) -- ++(0.5,0.1) node[right] {\footnotesize $\sum_{i \in \cI}\psi_i(t)$};
\draw[thick] (7,3) node[left] {\footnotesize} -- (7.3,3);
\draw[line width = 0.15,variable = \t, domain= 0 : 4, samples = 100] plot[smooth] (\t,{2.5 + 0.5*\t/7 + sin((0.5*pi+7.2*pi/(7.2-\t)) r)});
\draw[line width = 0.15,variable = \t, domain= 4 : 6, samples = 100] plot[smooth] (\t,{2*\t^3/27 - 11*\t^2/9 + 56*\t/9 - 164/27 - 1.5 + 0.5*\t/7 + sin((0.5*pi+7.2*pi/(7.2-\t)) r)});
\draw[line width = 0.1,variable = \t, domain= 6 : 6.61, samples = 100] plot[smooth] (\t,{2*\t^3/27 - 11*\t^2/9 + 56*\t/9 - 164/27 - 1.5 + 0.5*\t/7 + sin((0.5*pi+7.2*pi/(7.2-\t)) r)});
\draw[line width = 0.06,variable = \t, domain= 6.6 : 6.8, samples = 200] plot (\t,{2*\t^3/27 - 11*\t^2/9 + 56*\t/9 - 164/27 - 1.5 + 0.5*\t/7 + sin((0.5*pi+7.2*pi/(7.2-\t)) r)});
\draw[line width = 0.03,variable = \t, domain= 6.8 : 7, samples = 300] plot (\t,{2*\t^3/27 - 11*\t^2/9 + 56*\t/9 - 164/27 - 1.5 + 0.5*\t/7 + sin((0.5*pi+7.2*pi/(7.2-\t)) r)});

\draw (2.2,2.4698) -- ++(0.5,-0.1) node[right] {\footnotesize $W(t)$};
\foreach \k/\t in {0/1.4, 1/3, 2/4.5, 3/5.3, 4/6} 
{
   \fill[black] (\t,0) circle[radius=0.03];
   \draw (\t,0) node[below] {\footnotesize $\tau_{\k}$};
}
\draw (0,1.9756) -- (4,1.9756);
\draw (7,1.4817) -- (7.3,1.4817);
\draw[line width = 0.4, variable = \t, domain= 4 : 7, samples = 180]
plot[smooth] (\t, {2*0.4939*\t^3/27 - 11*0.4939*\t^2/9 + 56*0.4939*\t/9 - 164*0.4939/27});
\draw (2,1.9756) -- ++(0.1,-0.6) node[right] {\footnotesize $(1 - \eps_q)\sum_{i \in \cI} \psi_i(t)$};
\draw (0,2.9880) -- (4,2.9880);
\draw (7,2.2410) -- (7.3,2.2410);
\draw[line width = 0.4, variable = \t, domain= 4 : 7, samples = 180]
plot[smooth] (\t, {2*0.7470*\t^3/27 - 11*0.7470*\t^2/9 + 56*0.7470*\t/9 - 164*0.7470/27});
\draw (0.1,2.9880) -- ++(0.1,0.6) node[right] {\footnotesize $\left(1 - \frac{\eps_q}{2}\right)\sum_{i \in \cI} \psi_i(t)$};

\fill[red] (5.3,3.2671) circle [radius=0.03];
\draw[red,thin,dashed] (5.3,3.2671) -- (5.3,0) (5.3,-0.35) node[below] {\footnotesize \rotatebox{90}{$=$}} (5.3,-0.65) node[below] {\footnotesize $\tau_{k_q}$};

\fill[red] (4.96928,1.8543) circle [radius=0.03] (4.96928,0) circle [radius=0.03];
\draw[red,thin,dashed] (4.96928,1.8543) -- (4.96928,0);
\draw[red,thin] (4.96928,0) -- ++(-0.35,-0.7) node[anchor=north,inner sep =0]  {\footnotesize $s_q$};

\end{tikzpicture}
\caption{Illustration of the choice of the sequence $\{s_q\}_{q\in\N}$ based on $\{\tau_k\}_{k\in\N}$.}\label{fig:s^n+1_p}
\end{figure}

By \eqref{eq:jg1} and \eqref{eq:jg2}, the sequence $\{s_q\}$ is strictly increasing and  $\lim_{q \to \infty} s_q = \omega$.
Moreover, since $\lim_{q \to \infty} W(s_q)/\sum_{i \in \cI} \psi_i(s_q) = 1$,
\begin{equation}\label{eq:a1}
\lim_{q \to \infty} \frac{\nu_i(s_q)}{\psi_i(s_q)} = 1, \quad \forall i \in \cI.
\end{equation}
In addition, from \eqref{eq:jg1} and \eqref{eq:jg2}, it follows that the difference $W(s) - (1-\eps_q) \sum_{i\in\cI}\psi_i(s)$ cannot decrease at $s=s_q<\tau_{k_q}$, hence by Assumption \ref{assum:funnel}
\begin{equation}\label{eq:key}
\dot W(s_q) \geq 
(1-\eps_q)\sum_{i \in \cI} \dot \psi_i(s_q) \ge -N \theta_\psi, \quad \forall q \in \N.
\end{equation}

On the other hand, if we compute $\dot W$, then we have
\[\begin{aligned}
\dot W(t) &= \sum_{i \in \cI} \sum_{j \in \cN} \alpha_{ij} (f_j(t,x_j(t)) - f_i(t,x_i(t))) \\
&\quad + \sum_{i \in \cI} \sum_{j \in \cN} \alpha_{ij} ( \mu_j(t) - \mu_i(t)),
\end{aligned}
\]
where $\mu_k(t) := \mu_k(\nu_k(t)/\psi_k(t))$, $k\in\cN$, for simplicity.
We can bound the first sum by $M_0 := \sum_{i\in\cN} \sum_{j\in\cN} \alpha_{ij} M_f$, where the constant $M_f$ is such that 
\begin{align}\label{eq:meas_het}
      |f_j(t,x_j(t)) - f_i(t,x_i(t))| \le M_f, \quad \forall t \in [t_0,\omega), 
\end{align}
whose existence follows from Lemma~\ref{lem:finite} and Assumption~\ref{assum:fi} because $\omega$ is finite. 
Invoking \eqref{eq:sum_sum=0} for the index set $\cI$, we therefore have that
\begin{equation}\label{eq:key2}
\dot W(t) \le M_0 + \sum_{i \in \cI} \sum_{j \in \cN\backslash\cI} \alpha_{ij} ( \mu_j(t) - \mu_i(t) ).
\end{equation}
Let $\cJ := \cN \backslash \cI$ (which is non-empty).
Then, \eqref{eq:key} and \eqref{eq:key2} yield
\begin{align*}
\sum_{i \in \cI, j \in \cJ} \alpha_{ij} \mu_j(s_q) \geq\!\!\! \sum_{i \in \cI, j \in \cJ} \!\!\!\alpha_{ij} \mu_i(s_q) - M_0 - N\theta_\psi =: M_q.
\end{align*}
By the connectivity of the graph (Assumption \ref{assum:graph}), at least one $\alpha_{ij}$, where $i \in \cI$ and $j \in \cJ$, is positive.
Thus, it follows from \eqref{eq:a1} that $M_q \to \infty$ as $q \to \infty$.
Since 
\[
    \sum_{i \in \cI, j \in \cJ} \alpha_{ij} \mu_j(s_q) \le |\cI| \bar \alpha \sum_{j \in \cJ} \max\left\{\mu_j(s_q), 0\right\}
\]
where $\bar \alpha := \max_{i,j \in \cN} \alpha_{ij} > 0$, we have
\begin{equation}
\sum_{j \in \cJ} \max\left\{\mu_j(s_q), 0\right\} \ge \frac{M_q}{|\cI| \bar \alpha}.
\end{equation}
Therefore, for each sufficiently large $q$, there is an index $j_q \in \cJ$ such that $\mu_{j_q}(s_q) \ge M_q/(|\cJ| |\cI| \bar \alpha)$; hence
\[
    \mu_{j_q}\left( \frac{\nu_{j_q}(s_q)}{\psi_{j_q}(s_q)} \right) \to \infty, \quad \text{i.e.} \quad \frac{\nu_{j_q}(s_q)}{\psi_{j_q}(s_q)} \to 1.
\]
Since $\cJ$ is a finite set, there is a subsequence $\{\bar \tau_k\} = \{s_{q_k}\}$ such that $j^* = j_{q_k}\in\cJ$ and $\frac{\nu_{j^*}(\bar{\tau}_k)}{\psi_{j^*}(\bar{\tau}_k)}\to 1$.
Consequently,
\[
   \cI_+(\{\tau_k\}) \overset{\eqref{eq:a1}}{\subseteq} \cI_+(\{s_q\}) \subseteq \cI_+(\{\bar{\tau}_k\}).
\]
By construction, $j^*\in \cI_+(\{\bar{\tau}_k\}) \setminus \cI_+(\{\tau_k\})$ and we can conclude \eqref{eq:inc} as desired.\QEDB
\end{pf}

\begin{rem1}\label{rem:finite} 
(finite-time synchronization) We want to note that, in theory, finite-time synchronization (for a given $T>0$, $\lim_{t \to t_0 + T}|x_i(t) - x_j(t)| = 0$, for all $i,j \in \cN$) can also be achieved by the proposed method.
For this, take $\psi_i(t)$, for all $i \in \cN$, such that $\psi_i(t) > 0$ for $t \in [t_0, t_0 + T)$ and $\psi_i(t_0 + T) = 0$.
Then, the proof of Theorem \ref{thm:funnel} still holds with $\infty$ replaced by $t_0+T$.
In this case, the proposed coupling law cannot be used after the time $t_0+T$.
Another discontinuous coupling, such as \citep{coraggio20cdc}, may need to be employed in order to maintain the synchronization after $t_0+T$. 
\QEDD
\end{rem1}

\begin{rem1}\label{rem:pseudoglobal}
(pseudo-global property) The assumption of Theorem \ref{thm:funnel} requires boundedness of the initial conditions in the sense that $|\nu_i(t_0)|<\psi_i(t_0)$.
As a matter of fact, we can trivially satisfy this condition by taking $\psi_i(t) = 1/(t-t_0)$ so that $\psi_i(t_0)=\infty$, and adapt the proof of Theorem \ref{thm:funnel} to be valid in this case.
Anyway, we note that each agent can pick a sufficiently large individual gain $\psi_i$ and hence considering an initially infinite funnel is not necessary in most cases. \QEDD
\end{rem1}

\begin{rem1}(high order case)
The proof technique used for Theorem~\ref{thm:funnel} can easily be extended to high order, fully actuated agents given by
\begin{align*}
\dot{\boldsymbol{x}}_i(t) &= F_i(t, \boldsymbol{x}_i(t)) + \boldsymbol{u}_i(t, \boldsymbol{\nu}_i(t)) \in \mathbb{R}^n, \\ 
\boldsymbol{\nu}_i(t) &= {\rm col}(\nu_i^1(t), \dots, \nu_i^n(t)) = \sum_{j \in \mathcal{N}_i}\alpha_{ij}(\boldsymbol{x}_j(t) - \boldsymbol{x}_i(t)).
\end{align*}
In this case, the multi-dimensional funnel coupling can be chosen as an \emph{element-wise} type:
\begin{align*}
\boldsymbol{u}_i(t, \boldsymbol{\nu}_i) &:= {\rm col}\left(\mu_i\!\left(\frac{\nu_i^1}{\psi_i(t)}\right), \dots, \mu_i\!\left(\frac{\nu_i^n}{\psi_i(t)}\right)\!\right),
\end{align*}
or a \emph{maximum gain} type:
\begin{align*}
\boldsymbol{u}_i(t, \boldsymbol{\nu}_i) &:= \boldsymbol{\mu}_i\left(\frac{\boldsymbol{\nu}_i}{\psi_i(t)}\right) = \gamma_i\left(\frac{|\boldsymbol{\nu}_i|_\infty}{\psi_i(t)}\right)\frac{\boldsymbol{\nu}_i}{\psi_i(t)}.
\end{align*}
More interesting case is the underactuated case, which is an ongoing research.
\QEDD
\end{rem1}

So far we have seen that Theorem \ref{thm:funnel} guarantees that the diffusive term resides inside the funnel. The next theorem ensures that the control action remains uniformly bounded (even when the funnel boundaries $\psi_i(t)$ converges to zero) under mild additional assumptions.

\begin{thm1}\label{cor:bound_gain}
In addition to the assumptions of Theorem~\ref{thm:funnel}, assume that one of the following conditions hold.
\begin{enumerate}[(a)]
\item $f_i(t, x) \equiv F(t, x) + g_i(t, x)$ where $F(t, x)$ is globally Lipschitz with respect to $x$ uniformly in $t$ and there exists $\overline{M}_{g}$ such that $|g_i(t, x)| \le \overline{M}_{g}$ for all $i \in \cN$, $t \ge t_0$, and $x \in \mathbb{R}$.
\item There exists $\overline{M}_{x}$ such that $|x_i(t)| \le \overline{M}_{x}$ for all $i \in \cN$ and $t \ge t_0$.
\end{enumerate}
Then the input $u_i(t, \nu_i(t)) = \mu_i(\nu_i(t)/\psi_{i}(t))$ is bounded on $[t_0, \infty)$, i.e., there exists $\overline{M}_u > 0$ such that for all $t \in [t_0, \infty)$ and $i \in \cN$, we have $|u_i(t, \nu_i(t))| \le \overline{M}_u$. \QEDD
\end{thm1}

\begin{pf}
Note that the proof of Theorem \ref{thm:funnel} is still valid for the case $\omega = \infty$ as long as the condition \eqref{eq:meas_het} holds with $\omega = \infty$.
If this is the case, then there is no sequence $\{\tau_k\}$ for $\omega = \infty$ that makes the index sets $\cI_+(\{\tau_k\})$ and $\cI_-(\{\tau_k\})$ (in the proof of Theorem \ref{thm:funnel}) non-empty, and thus, there exists $\delta > 0$ such that $|\nu_i(t)/\psi_i(t)| < 1 - \delta$, for all $t \ge t_0$ and $i \in \mathcal{N}$ so that the claim follows.
Now, it can be seen that condition (a) ensures \eqref{eq:meas_het} with $\omega = \infty$, because
\begin{align*}
&|f_j(t, x_j(t)) - f_i(t, x_i(t))| \\
&\le |F(t, x_j(t)) - F(t, x_i(t))| + |g_j(t, x_j)| + |g_i(t, x_i)| \\
&\le \overline{L}|x_j(t) - x_i(t)| + 2\overline{M}_g \le 2\overline{L}\sqrt{N} \overline{\psi}/\lambda_2 + 2\overline{M}_g
\end{align*}
where $\overline{L}$ is the Lipschitz constant of $F$ and the third inequality follows from~\eqref{eq:a6}.
Condition (b) also guarantees condition \eqref{eq:meas_het} with $\omega = \infty$.
\QEDB
\end{pf}

\begin{rem1}
It is interesting to note that uniform boundedness of the vector fields is actually not required for Theorem~\ref{thm:funnel} and Theorem~\ref{cor:bound_gain}~(a).
In fact, we only need boundedness of $f_i$ on each compact subset of $[t_0, \infty) \times \R$.
\QEDD
\end{rem1}

We emphasize that the input remains bounded even if the state is unbounded as long as condition (a) of Theorem \ref{cor:bound_gain} holds (because the condition allows unbounded solution to the homogeneous part of the node dynamics $\dot{\bar{x}} = F(t, \bar{x})$).
This property is useful when one considers synchronization of unstable systems.
Also, the term $g_i$ can represent a perturbation of the state; even when the state $x_i$ is perturbed as $x_i + \tilde x_i$, the input can remain bounded if $g_i(t) = \dot {\tilde x}_i(t)$ satisfies condition (a).

We also emphasize that there are cases where condition (b) is guaranteed {\it a priori} before analyzing the effect of coupling inputs.
For example, if all node dynamics $\dot x_i = f_i(t,x_i)$, for all $i \in \cN$, are contractive (i.e., there exists $c_i>0$ such that $(\partial f_i /\partial x)(t, x_i) \le -c_i$, for all $t \ge t_0$ and $x_i \in \mathbb{R}$), then two dynamics of \eqref{eq:uplowdyn} have the same property almost everywhere and one can show the boundedness of all $x_i$'s using an inequality similar to \eqref{eq:dplus}.

\begin{rem1} (extension to conventional funnel control)
In the discussions so far, the performance function $\psi_i$ can converge to zero as time tends to infinity without a positive lower bound.
In fact, even if $\psi_i$ goes to zero, the ratio $\nu_i(t)/\psi_i(t)$ remains within a compact interval in $(-1,1)$ so that the input remains bounded.
The same idea can be applied for extending the conventional funnel controls, where the performance function has non-zero lower bounds and thus only practical tracking is guaranteed.
The readers are referred to \citep{jglee19cdc} for more on asymptotic tracking by funnel control with bounded inputs, which utilized the new funnel gain $\gamma_i(|\nu_i|/\psi_i(t))$ proposed in this paper.
\QEDD
\end{rem1}

\section{Emergent behavior under funnel coupling}\label{sec:Eb}

In Section~\ref{sec:MR} the system \eqref{eq:eachdyn} is proven to achieve (practical) synchronization by the funnel coupling law \eqref{eq:funnel_coup}, in the sense that, for all $i$ and $j$,
$$\limsup_{t\to\infty}|x_i(t)-x_j(t)| \le \frac{2\sqrt{N}}{\lambda_2} \limsup_{t\to\infty} \max_k \psi_k(t),$$ 
when the right-hand side is small or zero (see \eqref{eq:a6}).
In this section, we answer the question: when (practical) synchronization is achieved, what is the behavior of the agents?

We will show that, if (practical) synchronization is achieved, each agent behaves similar to the single scalar {\em emergent dynamics} given as
\begin{align}\label{eq:emer}
\dot{\xi} = h_\mu^{\psi}(t,f_1(t, \xi), \dots, f_N(t, \xi)) =: f_{\text{em}}(t, \xi)
\end{align}
with suitably chosen initial value, when the emergent dynamics is stable in a certain sense.
Here, the function~$h_\mu^{\psi}$ that maps ${\rm col}(t, f_1, \dots, f_N) \in \mathbb{R}^{N+1}$ to $h=h_\mu^{\psi}(t, f_1, \dots, f_N) \in \mathbb{R}$ is defined as the unique solution of the following algebraic equation\footnote{Note that, in \eqref{eq:alg}, $f_i$ is not the vector field of agent $i$ but just an arbitrary scalar argument of the function $H$.}:
\begin{align}\label{eq:alg}
H(h,t,f_1,\ldots,f_N):=\sum_{i=1}^N\psi_i(t) \mu_i^{-1}\left(h - f_i\right) = 0
\end{align}
where $\mu_i^{-1}:\R\to(-1,1)$ is well defined since $\mu_i:(-1,1)\to\R, s\mapsto \gamma_i(|s|)s$ is strictly increasing and surjective due to Assumption~\ref{assum:funnel}.
An intuition behind this equation is that, if all the states $x_i$ are synchronized to $\xi$, the time derivative $\dot x_i$ also should be the same as $\dot \xi$ across the agents.
This means that the difference in the vector field $f_i(t,\xi)$ across the agents should be compensated by individual $\mu_i$, so that $\dot \xi = \dot x_i = f_i(t,\xi) + \mu_i(\nu_i/\psi_i(t)) = f_{\text{em}}(t,\xi)$, for all $i$.
Recalling that $\sum_{i=1}^N \nu_i \equiv 0$ by construction, \eqref{eq:alg} follows.

\begin{lem1}\label{lem:alg}
Under Assumption \ref{assum:funnel}, there is a unique solution $h_\mu^{\psi} \in \mathbb{R}$ to \eqref{eq:alg} for each $(t, f_1, \dots, f_N)$.
Moreover, the solution satisfies
\begin{equation}\label{eq:hminmax}
\min_i f_i \le h_\mu^\psi \le \max_i f_i
\end{equation}
for any choice of $\psi_i$'s satisfying Assumption \ref{assum:funnel}.
If all $\mu_i$'s are continuously differentiable, then the map $(t, f_1,\ldots,f_N)\mapsto  h_\mu^{\psi}(t, f_1, \dots, f_N)$ is continuously differentiable.
\QEDD
\end{lem1}

By consequence of Lemma~\ref{lem:alg}, $f_{\text{em}}(t, \xi)$ as defined in~\eqref{eq:emer} is measurable in $t$ and locally Lipschitz in $\xi$, which guarantees existence and uniqueness of solutions of~\eqref{eq:emer}.

\begin{pf}
From \eqref{eq:funnel_coup}, the functions $\mu_i$ are strictly increasing in the inverval $(-1,1)$.
Hence, $\mu_i^{-1}$ is continuous and strictly increasing over $\R$.
This implies that, for each $(t,f_1,\ldots,f_N) \in \R^{N+1}$, the map $h \mapsto H(h,t,f_1,\ldots,f_N)$ is strictly increasing.
Because $H(h,t,f_1,\ldots,f_N)$ is positive if $h > \max_i f_i$ and is negative if $h < \min_i f_i$, there is a unique solution $h$ to \eqref{eq:alg} between $\min_i f_i$ and $\max_i f_i$ for each $(t,f_1, \dots, f_N)$.
Continuous differentiability of $h^{\psi}_\mu$ follows from the Implicit Function Theorem because $\partial H(h,t,f_1,\ldots,f_N)/\partial h > 0$ which is well-defined because $\gamma_i(0)>0$ from Assumption \ref{assum:funnel}. \QEDB
\end{pf}

\begin{rem1}\label{rem:commonpsi}(time-invariant emergent dynamics)
If all the performance functions $\psi_i$ share the same function $\psi$ as $\psi_i(t) = r_i \psi(t)$ where $r_i > 0$ are constants, then $\psi(t)$ can be removed from \eqref{eq:alg} so that the emergent dynamics \eqref{eq:emer} becomes a \emph{time-invariant} system which does not depend on $\psi$ but only on $r_i$. \QEDD
\end{rem1}

\begin{exam}\label{exam:1}
Consider $\mu_i(\eta) = \eta/(1 - |\eta|)$, for all $i \in \cN$, whose inverse is $\mu_i^{-1}(s) = s/(1 + |s|)$, and suppose that all $\psi_i$ are the same as $\psi$.
Then, \eqref{eq:alg} becomes
\[
    H(h,t,f_1,\ldots,f_N) = \sum_{i=1}^N \psi(t)\frac{h - f_i}{1 + |h - f_i|} \equiv 0.
\]
For each given $(t,f_1,\dots,f_N)$, this equation can be solved by the following procedure:
\begin{enumerate}[1.]
\item Find an index set $\{i_1, \dots, i_N\}$ such that $f_{i_j} \le f_{i_{j+1}}$ for all $j = 1, \dots, N-1$. Set $j = 1$.
\item Solve 
\begin{equation}\label{ex:1}
\sum_{k=1}^j \frac{h - f_{i_k}}{1 + h - f_{i_k}} + \sum_{k = j+1}^N \frac{h - f_{i_k}}{1 - h + f_{i_k}} = 0
\end{equation}
which is equivalent to finding roots of a polynomial in $h$ of order at most $N$.
\item If there is a root $h$ such that $f_{i_j} \le h \le f_{i_{j+1}}$ then return $h_\mu^\psi(t,f_1,\ldots,f_N) = h$.
\item If not, increase $j$ by $1$ and go back to Step 2.
\end{enumerate}
Well-posedness of this algorithm is guaranteed by the uniqueness of the solution in Lemma \ref{lem:alg}.
\QEDD
\end{exam}

\begin{exam}
If the funnel coupling law is given as
\[
    \mu_i(\eta) = \begin{cases} \ln(1/(1 - \eta)) &\mbox{ if } \eta \ge 0, \\ \ln(1 + \eta) &\mbox{ if } \eta < 0, \end{cases}
\]
then the inverse can be calculated as 
\[
\mu_i^{-1}(s) = \begin{cases} 1 - e^{-s} &\mbox{ if } s \ge 0, \\ -1 + e^s &\mbox{ if } s < 0. \end{cases}
\]
Proceeding similar to Example \ref{exam:1}, we get the same procedure as before, where \eqref{ex:1} is replaced with 
$$\sum_{k=1}^j \psi_{i_k}(t)(1 - e^{-h + f_{i_k}}) + \sum_{k= j+1}^N \psi_{i_k}(t)(-1 + e^{h - f_{i_k}}) = 0.$$
Note that this is simply a second order polynomial in terms of a new variable $\overline{h}=e^h$.
Uniqueness of the solution follows again from Lemma \ref{lem:alg}.
\QEDD
\end{exam}

\begin{rem1} (relation to \eqref{eq:blend})
The intuition, briefly discussed below \eqref{eq:alg}, seems universal and makes a connection between \eqref{eq:alg} and the averaged vector field studied in \citep{kim2016robustness,panteley2017synchronization,jglee18automatica}.
For example, the blended dynamics \eqref{eq:blend}, which emerges when a linear high-gain coupling law $u_i(t, \nu_i) = k\nu_i$ ($k \gg 1$) is used, can also be derived by \eqref{eq:alg}.
That is, imagine that $k$ is pushed towards infinity so that the states $x_i$ are synchronized to $\xi$ and the vector fields are also synchronized to $f_s$, i.e., $f_i(t, \xi) + k\nu_i = f_s$, for all $i \in \cN$.
Then, $f_s$ should satisfy $\sum_{i=1}^N (f_s - f_i(t, \xi)) \equiv 0$ by the algebraic constraint $\sum_{i=1}^N \nu_i \equiv 0$.
The solution $f_s$ is \eqref{eq:blend}.
\QEDD
\end{rem1}

Our argument that emergent dynamics~\eqref{eq:emer} approximates the synchronized behavior of the network~\eqref{eq:eachdyn} coupled via~\eqref{eq:funnel_coup} is based on the following assumption.

\begin{assum1}\label{assum:tech}
{\rm (emergent behavior)}
For all $i\in\cN$:
(a) $\partial f_i/\partial t : [t_0, \infty) \times \mathbb{R} \to \mathbb{R}$ is bounded on each compact subset of $\mathbb{R}$ uniformly in $t \in [t_0, \infty)$,
(b) $\gamma_i$ is continuously differentiable on $(0, 1)$, 
(c) there exists $\lambda_\psi > 0$ such that $|\dot{\psi}_i(t)| \le \lambda_\psi \psi_i(t)$, for all $t \ge t_0$,
and (d) there exists $r_\psi > 0$ such that $\psi_i(t) \le r_\psi \min_j\psi_j(t)$, for all $t \ge t_0$.
\QEDD
\end{assum1}

Note that conditions (c) and (d) of Assumption \ref{assum:tech} is not a restriction if there exists $c$ such that $0 < c \le \psi_i(t) \le \overline{\psi}$, for all $t \ge t_0$ and $i \in \cN$.

\begin{thm1}\label{thm:main2}
Under Assumptions \ref{assum:graph}--\ref{assum:tech}, assume the following:
\begin{itemize}

\item There are (normalized) performance functions $\{\bar\psi_i\}$ satisfying Assumptions \ref{assum:funnel} and \ref{assum:tech} which are normalized\footnote{The condition $|\bar\psi_i(t)|\le 1$ does not restrict the class of performance functions because, from \eqref{eq:alg}, scaling all the performance functions with a same constant does not change the emergent dynamics \eqref{eq:emer}.} as $|\bar \psi_i(t)| \le 1$, for all $t \ge t_0$ and $i \in \cN$, under which the emergent dynamics \eqref{eq:emer} is contractive; that is, there exists $c>0$ such that
\begin{equation}\label{eq:contractive_emer}
\frac{\partial f_{\rm em}(t,\xi)}{\partial \xi} \le -c, \quad \forall t \ge t_0, \; \xi \in \R.
\end{equation}

\item The initial condition $\boldsymbol{x}(t_0) = {\rm col}(x_1(t_0),\dots,x_N(t_0))$ of the system \eqref{eq:eachdyn} belongs to a compact set $C_0 \subset \R^N$.

\end{itemize}
With $\{\bar\psi_i\}$, let $\{\psi_i\}_{t_1,\eps}$ be a (parametrized) set of performance functions, where $t_1 > t_0$ and $\eps > 0$, such that
\begin{enumerate}
\item there exists $d>0$ such that, for all $i \in \cN$,
$$\max_{\boldsymbol{x} \in C_0} |\nu_i| + d = \max_{\boldsymbol{x} \in C_0} \left|\sum_{j \in \cN_i} \alpha_{ij} (x_j - x_i)\right| + d < \psi_i(t_0)$$
\item $\psi_i(t) = \eps \bar \psi_i(t)$, for all $t \in [t_1,\infty)$ and $i \in \cN$.
\end{enumerate}
Then, for each $\eta > 0$ and $\tau > 0$, there exists $\eps^*>0$ such that 
$$|x_i(t) - \xi(t)| \le \eta, \quad \forall t \ge t_1 + \tau, \quad \forall i \in \cN$$
where
\begin{itemize}
\item $\boldsymbol{x}$ is the solution to \eqref{eq:eachdyn} and \eqref{eq:funnel_coup} from an initial condition $\boldsymbol{x}(t_0) \in C_0$, with any choice of $\{\psi_i\}_{t_1,\eps}$ such that $0 < \eps \le \eps^*$,
\item $\xi$ is the solution to\footnote{Note that the emergent dynamics \eqref{eq:emer} by $\{\bar\psi_i\}$ and by $\{\psi_i\}$ are the same after $t_1$.} \eqref{eq:emer} from the initial condition $\xi(t_1+\tau) = (1/N)\sum_{i=1}^N x_i(t_1+\tau)$. 
\end{itemize}

If, in addition, $\lim_{t \to \infty} \bar{\psi}_i(t) = 0$, for all $i \in \cN$, then we further have
$$\lim_{t\to \infty} |x_i(t) - \xi(t)| = 0, \quad \forall i \in \cN.$$
\QEDD
\end{thm1}

The proof is given in Appendix~\ref{app:proofmain2}.

We emphasize that stability of individual agents are not required as long as the emergent dynamics is stable as in \eqref{eq:contractive_emer}.

\begin{rem1}(point-wise convergence)
Note also that, according to Appendix~\ref{app:proofmain2}, even without the contractive assumption~\eqref{eq:contractive_emer}, we can show point-wise convergence; for each $t > t_1$, we have
$$\lim_{\eps \to 0} \mu_i\left(\frac{\nu_i(t)}{\psi_i(t)}\right) = f_{\rm em}(t, x_i) - f_i(t, x_i), \quad i \in \cN,$$ 
which verifies our intuition that the $\mu_i$ term compensates the heterogeneity to yield the emergent dynamics~\eqref{eq:emer}.
\QEDD
\end{rem1}

\begin{rem1}
We note that it is hard to exactly identify the value $(1/N)\sum_{i=1}^Nx_i(t_1 + \tau)$, which depends on the network topology and the performance functions $\psi_i$ at $[t_0, t_1)$.\footnote{We want to emphasize that such a characterization is important only when we are interested in the approximation of the transient behavior.}
However, we can still ensure a reasonable estimate, because we can show that for any $\eta > 0$ there exists $t_1 + \tau > t_0$ which is sufficiently close to $t_0$ such that $\min_i x_{i}(t_0) - \eta \le (1/N)\sum_{i=1}^N x_{i}(t_1 + \tau) \le \max_i x_{i}(t_0) + \eta$ according to the arguments in the proof of Lemma~\ref{lem:finite}.
With this estimate, we can, for instance, make a transient error arbitrary small after an arbitrarily short time by making the stability of the emergent dynamics sufficiently strong.
Finally, we conjecture that the limit 
\[
   \lim_{t_1 +\tau \to t_0} \lim_{\eps \to 0}\frac{1}{N}\sum_{i=1}^N x_{i}(t_1+\tau)
\]
equals a weighted median of a collection $\chi$ of the initial values $x_i(t_0)$ with the weights $\psi_i(t_0)$, defined as a real number that belongs to the set $\cM_\chi$ defined in Section~\ref{subsec:app_dms}. 
We refer to Appendix~\ref{app:conj} for details.\QEDD
\end{rem1}

For further utility, we also note the following theorem.

\begin{thm1}\label{thm:asymp}
In addition to Assumptions \ref{assum:graph}--\ref{assum:tech}, assume that $|\nu_i(t_0)| < \psi_i(t_0)$ and $\lim_{t\to \infty} \psi_i(t) = 0$, for all $i \in \cN$, and that the emergent dynamics \eqref{eq:emer} is contractive.
If the solution $x_i(t)$ of \eqref{eq:eachdyn} with \eqref{eq:funnel_coup}, $i \in \cN$, is uniformly bounded, i.e., there exists $M_x$ such that $|x_i(t)| \le M_x$, for all $t \in [t_0, \infty)$ and $i \in \mathcal{N}$, then the steady-state behavior of the network follows that of the emergent dynamics, i.e., 
\[  \lim_{t\to \infty} |x_i(t) - \xi(t)| = 0, \quad i \in \mathcal{N},   \]
where $\xi(\cdot)$ is the solution of the emergent dynamics \eqref{eq:emer} with some initial condition $\xi(t_0) \in \mathbb{R}$.\footnote{Note that the initial condition $\xi(t_0)$ is irrelevant in the statement of Theorem~\ref{thm:asymp}, as any two trajectories of the emergent dynamics~\eqref{eq:emer} asymptotically converge to each other by the assumption that \eqref{eq:emer} is contractive.}\QEDD
\end{thm1}

\begin{pf}
By Theorem~\ref{cor:bound_gain}, there exists $\delta > 0$ such that $|\nu_i(t)/\psi_i(t)| \le 1-\delta$ for all $t \ge t_0$ and $i \in \cN$.
Then, by Lemma~\ref{lem:vu} and Appendix~\ref{app:asymp}, the proof concludes. \QEDB
\end{pf}

Now, given the characterization of the emergent dynamics~\eqref{eq:emer}, and given the analysis which shows that heterogeneous agents~\eqref{eq:eachdyn} under node-wise funnel coupling~\eqref{eq:funnel_coup} behaves accordingly with the emergent dynamics when the performance function is sufficiently narrow, we can, for instance, construct a heterogeneous network achieving a specific purpose as noted in the Introduction, if the emergent dynamics is contractive.
Note that under the assumption that all agents use the same funnel $\psi_i=\psi$ then the emergent dynamics~\eqref{eq:emer} only depend on the individual vector field $f_i$ and the coupling function $\mu_i$ for all $i \in \cN$, and thus, can be designed prior without knowing the performance function and the network topology.
This scheme of constructing a network with the desired collective behavior is first introduced in~\citep{jglee18automatica} and has many interesting applications.
Since the blended dynamics~\eqref{eq:blend} introduced in~\citep{jglee18automatica} (which corresponds to the emergent dynamics in this paper) takes clearly different form to the emergent dynamics~\eqref{eq:emer}, a new application might occur.
In fact, for any collection of coupling functions $\mu_i$, the function $h_\mu^\psi$ can never be linear, i.e., for each $t \ge t_0$ and ${\rm col}(a_1, \dots, a_N) \in \mathbb{R}^N$ there exists ${\rm col}(f_1, \dots, f_N) \in \mathbb{R}^N$ such that $h_\mu^\psi(t, f_1, \dots, f_N) \neq \sum_{i=1}^N a_if_i$.
In this regard, we further inspect the properties of the emergent dynamics~\eqref{eq:emer}, especially the properties of the function $h_\mu^\psi$ in the following section.

\section{Discussions on the emergent dynamics}\label{sec:disc}

\subsection{Numerical integration}\label{subsec:simul}

If the vector fields $f_i$ are differentiable, then the solution $\xi$ of the emergent dynamics \eqref{eq:emer} can be numerically obtained by the fact that the time derivative of $h_\mu^\psi(t,f_1,\dots,f_N)$, which is the solution to \eqref{eq:alg}, is again a function of known quantities like $\psi_i$, $\mu_i$, and $f_i$.
In particular, by invoking the Implicit Function Theorem to \eqref{eq:alg}, we have
\begin{align*}
\frac{\partial h_\mu^{\psi}}{\partial f_i}(t,f_1, \dots, f_N) &= \frac{\psi_i(t)(\mu_i^{-1})'(h^{\psi}_\mu - f_i)}{\sum_{j=1}^N \psi_j(t)(\mu_j^{-1})'(h^{\psi}_\mu - f_j)} \\
\text{and} \;\; 
\frac{\partial h_\mu^\psi}{\partial t}(t, f_1, \dots, f_N) &= - \frac{\sum_{j=1}^N \dot{\psi}_j(t)\mu_j^{-1}(h_\mu^\psi - f_j)}{\sum_{j=1}^N \psi_j(t)(\mu_j^{-1})'(h_\mu^\psi - f_j)}.
\end{align*}
Therefore, if we let $\chi = h_\mu^\psi(t, f_1(t, \xi), \dots, f_N(t, \xi))$, then
\begin{align}
\dot{\xi} &= \chi \nonumber \\
\dot{\chi} &=\! \frac{\sum_{j=1}^N\! \psi_j(t)(\mu_j^{-1})'(\chi\! -\! f_j(t, \xi))\!\left[ \frac{\partial f_j}{\partial t}(t, \xi) \!+\! \frac{\partial f_j}{\partial \xi}(t, \xi) \chi\right]}{\sum_{j=1}^N \psi_j(t)(\mu_j^{-1})'(\chi - f_j(t, \xi))} \nonumber \\
&\quad- \frac{\sum_{j=1}^N \dot{\psi}_j(t)\mu_j^{-1}(\chi - f_j(t, \xi))}{\sum_{j=1}^N\psi_j(t)(\mu_j^{-1})'(\chi - f_j(t, \xi))} 
\label{eq:chi}  \\
&=: g_\mu^\psi(t, \chi, \xi) \nonumber
\end{align}
with initial value $\xi(t_0)$ and $\chi(t_0) = h_\mu^\psi(t_0, f_1(t_0, \xi(t_0)), \allowbreak \dots, f_N(t_0, \xi(t_0)))$.

Note that when $\psi_i = r_i \psi$ (see Remark \ref{rem:commonpsi}) the partial derivative of $h_{\mu}^\psi$ with respect to time is zero and the dynamics is further simplified.

This is also robust with respect to numerical errors, when the emergent dynamics~\eqref{eq:emer} is contractive, in the sense that the linearization on the trajectory
\begin{align*}
\xi(t) &= \bar{\xi}(t), \quad \chi(t) = h_\mu^\psi(t, f_1(t, \bar{\xi}(t)), \dots, f_N(t, \bar{\xi}(t)))
\end{align*}
is exponentially stable, where $\bar{\xi}(\cdot)$ is the solution trajectory of~\eqref{eq:emer} with initial value $\bar{\xi}(t_0) = \xi(t_0)$.
In particular, if we transform the variable $\chi$ to $e_h := \chi - h_\mu^\psi(t, f_1(t, \xi), \dots, f_N(t, \xi)) = \chi - f_\text{em}(t, \xi)$, then we get
\begin{align*}
\dot{\xi} &= f_\text{em}(t, \xi) + e_h \\
\dot{e}_h &= g_\mu^\psi(t, e_h + f_\text{em}(t, \xi), \xi) - g_\mu^\psi(t, f_\text{em}(t, \xi), \xi).
\end{align*}
Therefore, its linearization reads as
\begin{align*}
\dot{\delta \xi} &= \frac{\partial f_\text{em}}{\partial \xi}(t, \bar{\xi}(t))\delta \xi + \delta e_h \\
\dot{\delta e}_h &= \frac{\partial g_\mu^\psi}{\partial \chi}(t, f_\text{em}(t, \bar{\xi}(t)), \bar{\xi}(t)) \delta e_h 
\end{align*}
where in the second equation all the terms associated with $\delta \xi$ cancels out.
Finally, noting that
\begin{align*}
    &\frac{\partial g_\mu^\psi}{\partial \chi}(t, f_\text{em}(t, \bar{\xi}(t)), \bar{\xi}(t)) = \frac{\partial f_\text{em}}{\partial \xi}(t, \bar{\xi}(t)) \\
    &\quad - \frac{d}{dt}\ln\left(\sum_{j=1}^N \psi_j(t)(\mu_j^{-1})'(f_\text{em}(t, \bar{\xi}(t)) - f_j(t, \bar{\xi}(t)))\right)
\end{align*}
the stability follows for performance functions having their convergence rate smaller than $c$, where $c$ is the contraction rate of the emergent dynamics.
This implies that we eventually have $\partial g_\mu^\psi/\partial \chi$ as negative.

\begin{exam}\label{exam:3}
Let us consider the network used in \citep{shim2015preliminary}, which consists of five agents of the form:
\begin{align*}
\dot{x}_i &= (-1 + \delta_i)x_i + c_i(t) + \mu_i(\nu_i/\psi(t)) \\
c_i(t) &= 10 \sin t \!+\! 10 m_i^1 \sin (0.1 t \!+\! \theta_i^1) \!+\! 10 m_i^2 \sin (10t \!+\! \theta_i^2) 
\end{align*}
where $\mu_i(s) = s/(1-|s|)$, $\psi(t) = 2 + 38 e^{-t}$, and $\delta_i$, $m_i^q$ and $\theta_i^q$ are some constants.
Since all the performance functions are the same, \eqref{eq:chi} simply becomes
\begin{align}\label{eq:diff}
\dot{\xi} &= \chi \\
\dot{\chi} &= \frac{\sum_{i=1}^N [(-1  + \delta_i)\chi + \dot{c}_i]/( 1  +  |\chi + (1 - \delta_i)\xi - c_i|)^2}{\sum_{i=1}^N 1/(1 + |\chi + (1 - \delta_i)\xi - c_i|)^2} \nonumber
\end{align}
and the corresponding simulation results are shown in Figure~\ref{fig:sim_emg_revisit}.

\begin{figure}[!ht]
\begin{center}
\includegraphics[width=.8\columnwidth]{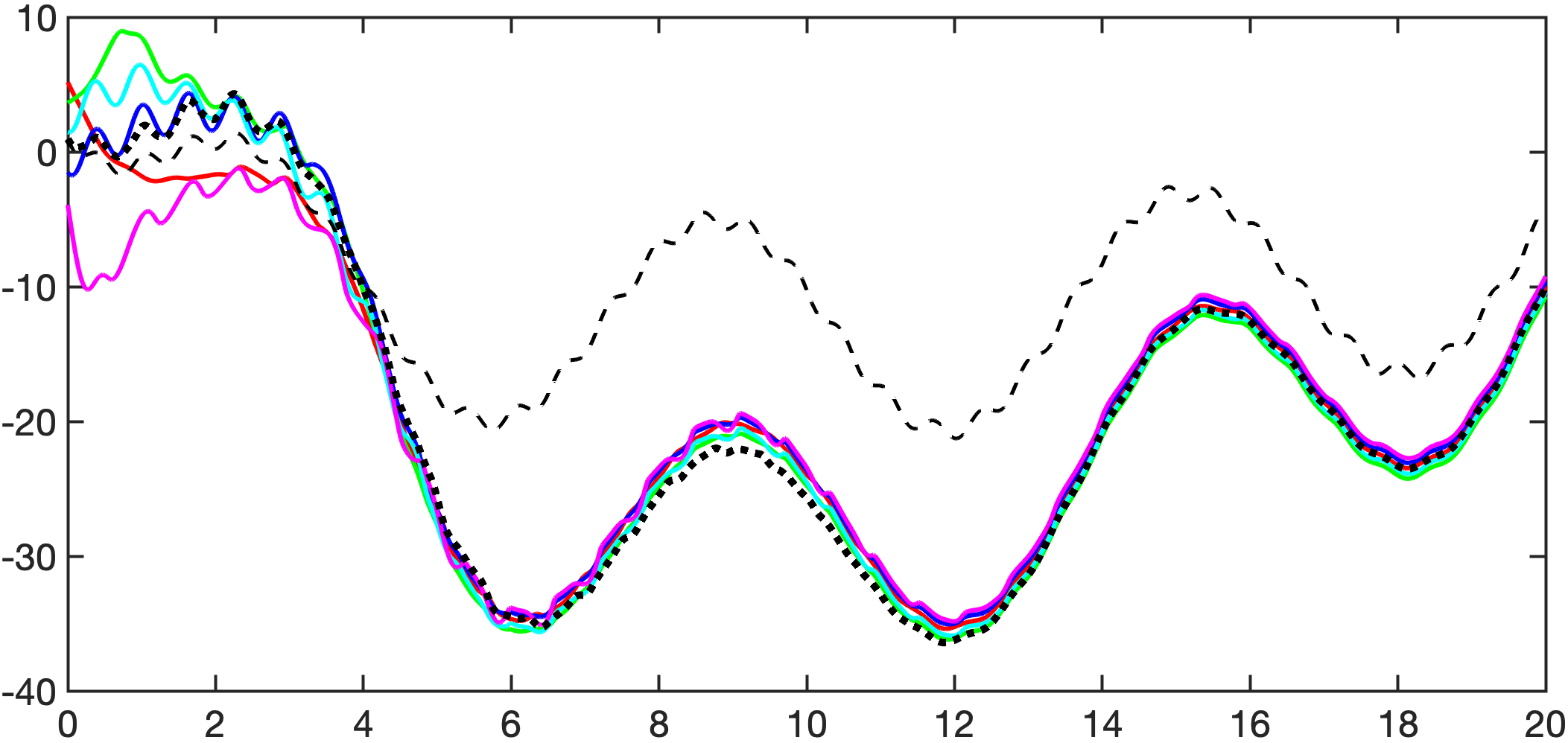}
\caption{Reproduced simulation from \citep{shim2015preliminary} with the solution to \eqref{eq:diff} plotted as a black dotted curve (behind the colored trajectories of each agent), which clearly predicts the synchronized behavior. The distinct black dashed curve is the solution of averaged dynamics $\dot s = (1/5)\sum_{i=1}^5 f_i(t,s)$ which is different from the synchronized behavior.}
\label{fig:sim_emg_revisit}
\end{center}
\end{figure}
\end{exam}

\subsection{Design of emergent dynamics}

Let us now discuss how to utilize the flexibility of choosing $\mu_i$ and $\psi_i$ towards achieving a desired emergent behavior.

\subsubsection{Electing a leader by designing $\psi_i$}\label{subsec:elect}

From the equation \eqref{eq:alg} it is seen that, if $\psi_{i^*}(t)$ is much larger than all others $\psi_j(t)$, then the solution $h$ tends to $f_{i^*}$.
This means that, in the situation when the agent $i^{*}$ wants to become the leader of the networked system, the function $\psi_{i^*}$ can be taken sufficiently large so that the emergent dynamics becomes similar to $\dot \xi = f_{i^*}(t,\xi)$.
Under Theorems \ref{thm:main2} or \ref{thm:asymp}, the collective behavior of the network becomes similar to the behavior of agent $i^*$.

The underlying intuition is that as $\psi_{i^*}(t) \to \infty$, the effect of the coupling law $u_{i^*}(t, \nu_{i^*}(t))$ becomes negligible.
In particular, when $(1-\eta)\psi_{i^*}(t)\ge \sum_{j \neq i^*}\psi_j(t)$ for all $t \ge t_0$ with some $\eta > 0$, we have, from the algebraic constraint $\sum_{i=1}^N \nu_i \equiv 0$, that
\begin{align*}
|\nu_{i^*}(t)| \le \sum_{j \neq i^*}|\nu_j(t)| < \sum_{j \neq i^*} \psi_j(t) <( 1-\eta)\psi_{i^*}(t),
\end{align*}
implying $|\mu_{i^*}(\nu_{i^*}(t)/\psi_{i^*}(t))| \le M_\eta := \mu_{i^*}(1-\eta) < \infty$ for all $t \ge t_0$, and thus, $V := |e_{i^*}| := |x_{i^*} - \bar{x}_{i^*}|$ satisfies
\begin{align*}
\dot{V} = \frac{e_{i^*}}{|e_{i^*}|}\dot{e}_{i^*} \le |f_{i^*}(t, e_{i^*} + \bar{x}_{i^*}) - f_{i^*}(t, \bar{x}_{i^*})| + M_\eta.
\end{align*}
where $\bar{x}_{i^*}$ is the solution of $\dot{\bar{x}}_{i^*}(t) = f_{i^*}(t, \bar{x}_{i^*}(t))$ with the initial condition $\bar{x}_{i^*}(t_0) = x_{i^*}(t_0)$.
Now, $V(t_0) = |e_{i^*}(t_0)| = 0$ gives for all $t \in [t_0, t_0 + \tau_\eta]$,
$$|e_{i^*}(t)| \!=\! V(t) \le \int_{t_0}^t e^{L(t - \tau)}M_\eta d\tau \le \frac{M_\eta}{L}e^{L\tau_\eta},$$
where $L$ is the Lipschitz constant of $f_{i^*}$ on the compact set $[-{M}_x, {M}_x]$, $T> 0$ is such that $\bar{x}_{i^*}(t) \in [-{M}_x, {M}_x]$ for all $t \in [t_0, t_0 + T]$, and $\tau_\eta \in (0, T)$ is such that $\bar{x}_{i^*}(t) + e_{i^*}(t) \in [-{M}_x, {M}_x]$ for all $t \in [t_0, t_0 + \tau_\eta]$.
Since, by its definition, we have $\lim_{\eta \to 1} {M}_\eta = 0$, and thus, $\lim_{\eta \to 1} \tau_\eta = T$, from which we get $\lim_{\eta \to 1} x_{i^*}(t) = \bar{x}_{i^*}(t)$ for all $t \in [t_0, t_0 + T)$.
Noting that the choice of ${M}_x$, hence $L$ and $T$, was arbitrary, this extends to $[t_0, \infty)$.

\begin{rem1}
We want to note that by the similar arguments as above, it can be proved that, when there exists $i^* \in \cN$ such that $(1 - \eta) \psi_{i^*} \ge \sum_{j \neq i^*} \psi_j $ with some $\eta > 0$, and when the dynamics $\dot{x} = f_{i^*}(t, x)$ is contractive, the solution trajectory is uniformly bounded on $[t_0, \infty)$, hence also the inputs.
This is because, $V := |e_{i^*}|$ now satisfies $\dot{V}  \le -c V + M_\eta$ with some $c > 0$, which gives the boundedness of $x_{i^*}$ from the boundedness of $\bar{x}_{i^*}$. \QEDD
\end{rem1}

\subsubsection{Effect of locally linear $\mu_i$}

If a particular behavior is desired for a group of heterogeneous multi-agent system, the behavior can be achieved by suitably designing the emergent dynamics and by Theorems \ref{thm:main2} or \ref{thm:asymp}.
And the design becomes easier if the emergent dynamics is simply a linear combination of individual node dynamics, like in \eqref{eq:blend}.
(See \citep{jglee18automatica} for a few design examples by \eqref{eq:blend}.)
Even with the nonlinear coupling $\mu_i$, this is possible if $\mu_i$, $i \in \cN$, are locally linear.
Suppose that \eqref{eq:blend} is stable and $\xi(t)$ of \eqref{eq:blend} remains in a certain compact set $[-M_x, M_x] \subset \mathbb{R}$.
Let $M_f := \sup_{i, t, |x| \le M_x} |f_i(t, x)|$ and take 
\[
   \mu_i(s) = \begin{cases} 4 M_f s, &\text{if } |s| < 0.5, \\ 4 M_f s + \frac{s}{1-|s|} - 4 s |s|, &\text{if } |s| \in [0.5, 1), \end{cases}
\]
for all $i \in \cN$.
Then, $\mu_i^{-1}$ is linear in the interval $[-2M_f,2M_f]$.
Now, let all the performance functions be identical as $\psi_i=\psi$.
Then, \eqref{eq:alg} becomes 
\[  0 = \sum_{i=1}^N \mu_i^{-1} (h_\mu^\psi - f_i(t, \xi)) = \sum_{i=1}^N\frac{h_\mu^\psi - f_i(t, \xi)}{4M_f}   \]
because $h_\mu^\psi-f_i(t,\xi) \in [-2M_f,2M_f]$ by \eqref{eq:hminmax}.
It is now clear that, for all $\xi \in [-M_x,M_x]$,
\[h_\mu^\psi(t, f_1(t, \xi), \dots, f_N(t, \xi)) = \frac1N \sum_{i=1}^N f_i(t, \xi).\]

\begin{rem1}
In \citep{jglee18automatica}, a linear coupling of the form $u_i = k \nu_i$ is used, which yielded several limitations.
The convergence is semi-global and practical, and the threshold for the gain $k$ depends on the global information such as network structure.
On the contrary, the proposed funnel coupling does not rely on such global information, leads to asymptotic convergence, and can be made pseudo-global as discussed in Remark \ref{rem:pseudoglobal}.
\QEDD
\end{rem1}

\begin{rem1}
In \citep{shim2015preliminary}, the coupling function $\mu(s) = \kappa s / (1-|s|)$ is used with an observation that, when the parameter $\kappa$ gets larger, the synchronized behavior gets closer to the behavior of \eqref{eq:blend}.
This observation can now be explained by the fact that $\mu^{-1}(u) = u/(\kappa+|u|)$ gets more linear in a local region when $\kappa$ gets larger. \QEDD
\end{rem1}

\subsubsection{Finding median agent by $\mu_i$}\label{subsec:app_dms}

For a collection $\mathcal{F} = \{ (f_i,\psi_i) : i \in \cN \}$ where $f_i$ is a number and $\psi_i > 0$ is a weight, the {\em weighted median} is defined as a number that belongs to the set
\[
   \cM_{\mathcal{F}} = \begin{cases} \{f_{s_j}\}, &\mbox{if } \exists j \in \cN, \, \sum_{k=1}^{j} \psi_{s_k} > \psi_{\text{half}} \\
&\,\,\quad\quad \text{ and }\sum_{k=1}^{j-1} \psi_{s_k}  < \psi_{\text{half}}, \\ 
[f_{s_j}, f_{s_{(j+1)}}], &\mbox{if } \exists j \in \cN, \, \sum_{k=1}^j \psi_{s_k}  =  \psi_{\text{half}},\end{cases}
\]
where $\psi_{\text{half}} := (1/2)\sum_{i=1}^N \psi_i$ and $\{s_k\}$ is the rearrangement of the sequence $\{1, \dots, N\}$ such that $f_{s_1} \le f_{s_2} \le \cdots \le f_{s_N}$.
Then, there are finitely many index sets $\mathcal{K} \subseteq \mathcal{N}$ such that $\sum_{i \in \mathcal{K}} \psi_i > \psi_\text{half}$.
Take $\delta > 0$ so that it holds for all such sets $\mathcal{K}$ that 
\begin{equation}\label{eq:delta}
\sum_{i \in \mathcal{K}} \psi_i \ge \left(\frac12 + \delta\right) \sum_{i \in \cN} \psi_i.
\end{equation} 
Now, for any $\eta > 0$ and $\eps$ such that $0 < \eps < 4\delta/(2\delta + 1)$, consider an equation
\begin{equation}\label{eq:virtualalg}
\sum_{i=1}^N \psi_i \mu_i^{-1}(h - f_i) = 0
\end{equation}
where $\mu_i^{-1}(s)$ is any function satisfying
\begin{align}\label{eq:beta_bound}
\begin{split}
\mu_i^{-1}(s) &\ge 1 - \eps, \quad s \ge \eta, \\ 
\mu_i^{-1}(s) &\le -1+\eps, \quad s \le -\eta. 
\end{split}
\end{align}
When both $\eps$ and $\eta$ are small, this function looks like a signum function.

\begin{lem1}\label{lem:med}
The solution $h$ to \eqref{eq:virtualalg} satisfies $| h |_{\cM_{\mathcal{F}}} \le \eta$. \QEDD
\end{lem1}

The proof is found in Appendix \ref{ssec:proof_lemma_median}.
Based on the lemma, the emergent dynamics \eqref{eq:emer} can be made arbitrarily close to a weighted median of individual vector fields $f_i$.
This is useful when, for example, one is interested in synchronization of a multi-agent system consisting of mostly identical agents but with a few outliers, and the effect of those outliers should be rejected.
(Refer to \citep{jglee19tac} to see how median operation can be used for rejecting malicious attack in multi-agent setting.)

\begin{exam}
Suppose that individual agent contains their own value $f_i^*$, and let us design a network that asymptotically finds a median of the data $\{f_i^*\}$.
By taking identical $\psi_i(t) = \psi(t)$, the weighted median $\cM_{\mathcal{F}^*}$ becomes the standard median, and thus, we can take $\delta = 1/(2N)$ and $\eps < 2/(N+1)$.
Let $\mu_i$, $i \in \cN$, satisfy \eqref{eq:beta_bound}, and let $\psi$ satisfy $\lim_{t\to\infty}\psi(t)=0$ and Assumptions \ref{assum:funnel} and \ref{assum:tech}.
Then, the proposed multi-agent system
\begin{align}\label{eq:dms}
\dot{x}_i(t) &= f_i^* - x_i(t) + \mu_i\left(\frac{\nu_i(t)}{\psi(t)}\right), \quad x_i(t_0) = x_i^0,
\end{align}
ensures asymptotic synchronization by Theorem \ref{thm:funnel}.
Moreover, since the solution $x_i(t)$ are uniformly bounded (which can be shown by a similar argument as the proof of Lemma \ref{lem:finite}), Theorem \ref{thm:asymp} ensures that the steady-state behavior of the network follows that of the contractive emergent dynamics
\begin{align*}
\dot{\xi}(t) &= h_\mu^\psi(f_1^* - \xi(t), \dots, f_N^* - \xi(t)) \\
&= h_\mu^\psi(f_1^*, \dots, f_N^*) - \xi(t)
\end{align*}
where $h_\mu^\psi(f_1^*,\ldots,f_N^*)$ is the solution to \eqref{eq:virtualalg}, because $\mu_i^{-1}(h-f_i) = \mu_i^{-1}((h-\xi) - (f_i-\xi))$.
Since $\xi(t)$ converges to the constant $h_\mu^\psi(f_1^*, \dots, f_N^*)$, which can be made arbitrarily close to a median of $\{f_i^*\}$ as shown in Lemma~\ref{lem:med}, the proposed scalar network finds a median with arbitrary precision.
Note that the design can be done in a fully decentralized manner, with the only prior agreement on $\eps$ and $\eta$, and that the median can be found without communicating the values $f_i^*$ to the neighbors, hence preserving privacy and increasing security.
\QEDD
\end{exam}

\section{Conclusion}\label{sec:con}

This paper introduces the funnel coupling law which guarantees synchronization for a heterogeneous multi-agent system under only mild assumptions.
Some sufficient conditions which guarantee boundedness of the inputs are also provided, and the analysis on the emergent collective behavior that appears as we enforce synchronization by the proposed funnel coupling law has been conducted.
In fact, the paper introduced emergent dynamics that can illustrate the synchronized behavior of the whole network, and from its nonlinear structure, some new applications have been discovered, e.g., distributed median solver.
Our future work is to extend our result to its vector counterpart, hence utilizing its interesting features, and to further derive useful applications.
Consideration of unknown input gain that may depend on time and state as in conventional funnel control is also of our future interest.

\bibliographystyle{pre-automatica}
\bibliography{root}

\appendix

\section{Proof of Theorem \ref{thm:main2}}\label{app:proofmain2}

For the proof, we define two new variables $x_s \in \R$ and $y \in \R^{N-1}$ as
\begin{align*}
x_s &:= \frac1N \sum_{i=1}^N x_i \\
y &:= -\frac{1}{\underline{\psi}(t)}\Lambda R^\top\!\!\begin{bmatrix} x_1 \\ \vdots \\ x_N\end{bmatrix} \! - \frac{1}{\underline{\psi}(t)}R^\top\!\!\begin{bmatrix} \psi_1(t)\mu_1^{-1}(f_{\text{em}}^s - f_1^s) \\ \vdots \\ \psi_N(t)\mu_N^{-1}(f_{\text{em}}^s - f_N^s)\end{bmatrix}
\end{align*}
where $\underline{\psi}(t) := \min_i \psi_i(t)$, the matrices $\Lambda$ and $R$ are defined around \eqref{eq:a6}, and $f_{\text{em}}^s$ and $f_i^s$ denote $f_{\text{em}}(t, x_s)$ and $f_i(t, x_s)$, respectively.
Then, it always holds that
\begin{equation}\label{eq:y}
\|y\| \le 2 r_\psi \sqrt{N}
\end{equation}
which can be seen from the facts that $\Lambda R^\top \boldsymbol{x} = (R^\top R)\Lambda R^\top \boldsymbol{x} = R^\top \mathcal{L} \boldsymbol{x} = -R^\top {\rm col}(\nu_1,\dots,\nu_N)$, $|\nu_i(t)| < \psi_i(t)$, $\psi_i(t)/\underline{\psi}(t) \le r_\psi$, $\|R\| = 1$, and $|\mu_i^{-1}(a)| \le 1$, for all $a$.
In addition, it can be seen that
\begin{align}\label{eq:omega_id}
\frac{\underline{\psi}(t)}{\psi_i(t)} R_i y = 
\frac{\nu_i(t)}{\psi_i(t)} - \mu_i^{-1}\left(f_{\text{em}}^s- f_i^s\right) 
\end{align}
because $R_iR^\top = e_i^\top - (1/N)1_N^\top$ where $e_i$ is the $i$-th elementary vector, and $\sum_{i=1}^N \psi_i(t)\mu_i^{-1}(f_{\rm em}^s - f_i^s) = 0$ by the definition of $f_{\rm em}$.
Now, we obtain
\begin{align}
\dot{x}_s &= f_{\text{em}}(t, x_s) + \frac{1}{N}\sum_{i=1}^N [f_i(t, x_i) - f_i(t, x_s)] \nonumber\\
&\quad + \frac{1}{N}\sum_{i=1}^N \left[\mu_i\!\left(\!\frac{\underline{\psi}(t)}{\psi_i(t)} R_i y + \mu_i^{-1}(f_{\text{em}}^s - f_i^s)\!\right) \right. \nonumber \\
&\quad\quad\quad\quad\quad\quad\quad\quad\quad \left.- \mu_i\left(\mu_i^{-1}(f_{\text{em}}^s  - f_i^s)\right)\right] \label{eq:deriv_x_s}
\end{align}
and
\begin{align}
&\dot{y} = \frac{\dot{\underline{\psi}}}{\underline{\psi}^2}\Lambda R^\top\begin{bmatrix} x_1 \\ \vdots \\ x_N\end{bmatrix} - \frac{1}{\underline{\psi}}\Lambda R^\top\! \begin{bmatrix} f_1(t, x_1) - f_1(t, x_s) \\ \vdots \\ f_N(t, x_N) - f_N(t, x_s)\end{bmatrix}\nonumber\\ 
&\hspace{-4mm} - \frac{1}{\underline{\psi}}\Lambda R^\top\!\! \begin{bmatrix} \mu_1\!\!\left(\frac{\underline{\psi}}{\psi_1} R_1 y \!+\! \mu_1^{-1}\!(f_{\text{em}}^s \!-\! f_1^s)\!\right) \!-\! \mu_1\!\left(\mu_1^{-1}\!(f_{\text{em}}^s \!-\! f_1^s)\!\right) \!\\ \vdots \\ \mu_N\!\!\left(\frac{\underline{\psi}}{\psi_N} R_N y \!+\! \mu_N^{-1}\!(f_{\text{em}}^s \!-\! f_N^s)\!\right) \!-\! \mu_N\!\left(\mu_N^{-1}\!(f_{\text{em}}^s \!-\! f_N^s)\!\right)\!\end{bmatrix} \nonumber\\ 
&\hspace{-2mm} - \!R^\top\!\!\begin{bmatrix}\frac{\underline{\psi}\dot{\psi}_1 - \dot{\underline{\psi}}\psi_1}{\underline{\psi}^2}\mu_1^{-1}(f_{\text{em}}^s \!-\! f_1^s) \\ \vdots \\ \frac{\underline{\psi}\dot{\psi}_N - \dot{\underline{\psi}}\psi_N}{\underline{\psi}^2}\mu_N^{-1}(f_{\text{em}}^s \!-\! f_N^s)\end{bmatrix} \!-\! R^\top\!\!\begin{bmatrix} \frac{\psi_1}{\underline{\psi}}\frac{d}{dt}\mu_1^{-1}\!(f_{\text{em}}^s \!-\! f_1^s) \\ \vdots \\ \frac{\psi_N}{\underline{\psi}}\frac{d}{dt}\mu_N^{-1}\!(f_{\text{em}}^s \!-\! f_N^s)\!\end{bmatrix} \label{eq:deriv_y}
\end{align}
(where we used $R^\top {\rm col}(f_{\text{em}}^s,\dots,f_{\text{em}}^s) = 0$).
Then, let us introduce two functions 
$$V(t) := |x_s(t) - \xi(t)|, \quad U(t) := \sqrt{y^\top(t) \Lambda^{-1} y(t)}$$
for which, $V(t_1+\tau) = 0$ from the assumption.
In the following, we will show that there is $\eps^*>0$ such that, when $0< \eps \le \eps^*$, we have that 
\begin{equation}\label{eq:pfgoal}
V(t) \le \frac{\eta}{2}, \qquad \forall t \ge t_1+\tau.
\end{equation}
This yields, with $\widehat{\psi}(t) := \max_i \psi_i(t) \le \eps$,
\begin{align}\label{eq:a.5}
\begin{split}
&|x_i(t) - \xi(t)| \le V(t) + |x_s(t) - x_i(t)| \\
&\quad \le V(t) + \|RR^\top\boldsymbol{x}(t)\|_\infty < \frac{\eta}{2} + \sqrt{N}\frac{\widehat{\psi}(t)}{\lambda_2} \le \eta
\end{split}
\end{align}
for all $t \ge t_1+\tau$ and $i \in \cN$, if $\eps^* \le \lambda_2\eta/(2\sqrt{N})$ (where we used $RR^\top = I-(1/N)1_N1_N^\top$ and \eqref{eq:a6}).

In order to obtain \eqref{eq:pfgoal}, we will analyze $V$ and $U$, and their time derivatives, for which a few bounds are useful.
First, there exists $M_0$ such that
\begin{equation}\label{eq:m0}
|x_i(t)| \le M_0, \qquad \forall t \in [t_0,t_1+\tau]
\end{equation}
which can be found by Lemma \ref{lem:finite} since $\boldsymbol{x}(t_0) \in C_0$.
We note that $M_0$ is independent of the choice of $\{\psi_i\}$ (see the proof of Lemma \ref{lem:finite}).
Now, note that the solution $\xi(t)$ of the emergent dynamics is uniformly bounded for $t \ge t_1+\tau$.
Indeed, it follows (from \eqref{eq:contractive_emer}) with $v(t) := \xi^2(t)/2$ that
\begin{align*}
\dot v &= \xi(f_{\rm em}(t,\xi) - f_{\rm em}(t,0)) + \xi f_{\rm em}(t,0) \\
&\le -c \xi^2 + |\xi| \sup_{t \ge t_1+\tau} |f_{\rm em}(t,0)| =: -c \xi^2 + |\xi| M_{\rm em} .
\end{align*}
Hence, $|\xi(t)| \le \max\{ |\xi(t_1+\tau)|, M_{\rm em}/c \}$, for all $t \ge t_1+\tau$, and since $|\xi(t_1+\tau)|=|x_s(t_1+\tau)| \le M_0$, we have $|\xi(t)| \le \max\{M_0,M_{\rm em}/c\} =: M_\xi$.
Here we make:

\noindent{\em Temporary assumption:} $|x_i(t)| \le M_x := M_\xi + \eta$, for all $t \ge t_0$.

This assumption trivially holds both for $t_0 \le t \le t_1+\tau$ by \eqref{eq:m0}, and for a certain amount of time after $t_1+\tau$.
The latter is because $|x_i(t_1+\tau)| \le |x_i(t_1+\tau)-\xi(t_1+\tau)| + |\xi(t_1+\tau)| \le \eta/2 + M_\xi = M_x - \eta/2$ (the second inequality is from \eqref{eq:a.5} with $\eps \le \eps^*$ since $V(t_1+\tau)=0$).
We will show that this period of time extends to infinity so that the temporary assumption turns out to be true.

Now, it follows from the temporary assumption that:
\begin{enumerate}[(i)]
\item there exist $\theta_f$, $L_f$, $M_f$ and $\delta_1>0$, such that for all $a \in [-M_x,M_x]$, $t \ge t_0$, and $i \in \cN$,
\begin{gather*}
\left| f_i(t,a) \right| \le M_f , \; 
\left| \frac{\partial f_i}{\partial t}(t,a) \right| \le \theta_f, \;
\left| \frac{\partial f_i}{\partial x}(t,a) \right| \le L_f, \\
\left|\mu_i^{-1}\left(f_{\text{em}}(t, a)- f_i(t, a)\right)\right| \le 1- 2\delta_1
\end{gather*}
in which, $\delta_1$ is independent of particular choice of $\{\bar\psi_i\}$ because $\min_i f_i(t,a) \le f_{\rm em}(t,a) \le \max_i f_i(t,a)$ by construction.

\item there exists $\delta_2 > 0$ such that
\begin{equation}\label{eq:nupsi}
\left|\frac{\nu_i(t)}{\psi_i(t)}\right| \le 1 - \delta_2, \quad \forall t \in [t_0, t_1 + \tau], \; i \in \cN
\end{equation}
with arbitrary performance functions $\psi_i$, which follows from the following argument.

Let
$$\omega_i := f_i(t, x_i) + \mu_i\left(\frac{\nu_i}{\psi_i(t)}\right) + \lambda_\psi x_i,$$
from which, we have
$$\dot{x}_i = -\lambda_\psi x_i + \omega_i.$$
Then, choose a time-varying index $J(t) \in \cN$ such that $\omega_{J(t)}(t) = \max_i \omega_i(t)$ and $\dot{\omega}_{J(t)}(t) \ge \dot{\omega}_k(t)$ for all $k \in \cN$ such that $\omega_k(t) = \max_i \omega_i(t)$.
The upper right Dini derivative of $\omega_{J(t)}(t)$ satisfies
\begin{align*}
&D^+ \omega_{J(t)} = \frac{\partial f_{J(t)}}{\partial t} + \frac{\partial f_{J(t)}}{\partial x}\left[\omega_{J(t)}- \lambda_\psi x_{J(t)}\right] \\
&\quad\quad + \mu'_{J(t)}\left(\frac{\nu_{J(t)}}{\psi_{J(t)}}\right) \cdot \left[-\frac{\dot{\psi}_{J(t)}}{\psi_{J(t)}} - \lambda_\psi\right]\frac{\nu_{J(t)}}{\psi_{J(t)}} \\
&\quad\quad +\mu'_{J(t)}\left(\frac{\nu_{J(t)}}{\psi_{J(t)}}\right) \cdot \frac{1}{\psi_{J(t)}}\sum_{j \in \cN_{J(t)}} \alpha_{J(t)j}[\omega_j - \omega_{J(t)}] \\
&\quad\quad + \lambda_\psi \left[\omega_{J(t)} - \lambda_\psi x_{J(t)}\right].
\end{align*}
By the definition of $J(t)$, the third term is non-positive if and only if $\nu_{J(t)} \ge 0$, and the fourth term is always non-positive.
Therefore, we can conclude that either
\begin{align*}
D^+\omega_{J(t)} &\le \left[\theta_f + \lambda_\psi L_f M_x + \lambda_\psi^2M_x\right] + \left[L_f + \lambda_\psi\right] |\omega_{J(t)}|
\end{align*}
when $\nu_{J(t)} \ge 0$, or from the definition of $\omega_i$,
$$\omega_{J(t)} \le f_{J(t)}(t, x_{J(t)}) + \lambda_\psi x_{J(t)} \le M_f + \lambda_\psi M_x$$
when $\nu_{J(t)}<0$.
By an analogous argument for the case when $\omega_{J(t)}(t) = \min_i \omega_i(t)$, we can thus find $M_\omega > 0$ such that
$$|\omega_i(t)| \le M_\omega, \quad \forall t \in [t_0, t_1 + \tau], \,\, i \in \cN.$$
Now, there exists $\delta_2 > 0$ such that for all $t \in [t_0, t_1+ \tau]$ and $i \in \cN$,
\begin{align*}
\left|\frac{\nu_i}{\psi_i(t)}\right| &\le \left|\mu_i^{-1}\left(\omega_i - f_i(t, x_i) - \lambda_\psi x_i\right)\right| \\
&\le \mu_i^{-1} \left(M_\omega + M_f + \lambda_\psi M_x\right) \le 1 - \delta_2.
\end{align*}

\item Let $\delta := \min\{\delta_1,\delta_2\}$. Then, there exists $L_\mu$ such that
$$\left|\mu_i'(a)\right| \le L_\mu, \quad \forall a \in [-1+\delta,1-\delta], \; i \in \cN.$$

\end{enumerate}

\begin{lem1}\label{lem:vu}
Under the temporary assumption, we have 
\begin{align}
\dot V &\le -cV + M_V \widehat \psi(t) + L_\mu \sqrt{\lambda_N} U \label{eq:V} \\
\dot U &\le -\left( \frac{\underline{\gamma}\lambda_2}{\widehat \psi(t)} - \lambda_\psi\right)  U + \frac{M_U \sqrt{\lambda_N}}{2} \label{eq:U}
\end{align}
where $M_V := L_f \sqrt{N}/\lambda_2$ and $M_U > 0$ (see \eqref{eq:mu2}), whenever
$$\left|\frac{\nu_i(t)}{\psi_i(t)}\right| \le 1 - \delta, \quad \forall i \in \cN.$$
\end{lem1}

\begin{pf}
See Appendix~\ref{app:Lemma}.
\end{pf}

Now, with 
$$\delta_\eta := \min\left\{ \frac{c\eta}{6L_\mu}, \delta, 3 r_\psi \right\}$$
let $\eps^* > 0$ be such that 
\begin{align*}
\eps^* &\le \min\left\{\frac{c\eta}{6M_V}, \frac{M_U\lambda_N}{2\underline{\gamma}\lambda_2\delta_\eta}, \frac{\underline{\gamma}\lambda_2}{2\lambda_\psi}, \frac{\underline{\gamma}\lambda_2 \delta_\eta}{2M_U\lambda_N}\right\} \\
\text{and} \quad \eps^* &\le \frac{ \underline{\gamma}\lambda_2\tau}{2\ln\left(\frac{4r_\psi}{\delta_\eta}\sqrt{\frac{N\lambda_N}{\lambda_2}}\right)}
\end{align*}
where $\underline{\gamma} := \min_i\gamma_i(0) > 0$.
Then, from \eqref{eq:y}, we have $U(t_1) \le 2 r_\psi \sqrt{N/\lambda_2}$.
Since $\widehat \psi(t) \le \eps \le \eps^*$ for $t \ge t_1$, it can be shown that inequality \eqref{eq:U} implies 
\begin{equation}\label{eq:Ulast}
U(t_1 + \tau) \le \frac{\delta_\eta}{\sqrt{\lambda_N}}.
\end{equation}
Indeed, we have from~\eqref{eq:U},
\begin{align*}
U(t_1 + \tau) &\le e^{-\lambda_\eps \tau}U(t_1) + \frac{M_U\sqrt{\lambda_N}}{2\lambda_\eps} 
\end{align*}
where $\lambda_\eps := \underline{\gamma}\lambda_2/(2\eps) \le \underline{\gamma}\lambda_2/\eps - \lambda_\psi$.

Now, we will show that the set 
$$U(t) \le \frac{\delta_\eta}{\sqrt{\lambda_N}}, \quad V(t) \le \frac{\eta}{2}$$
is positively invariant from $t = t_1 + \tau$, which concludes our proof.

For this purpose, note first that $U(t) \le \delta_\eta/\sqrt{\lambda_N}$ implies $\|y(t)\| \le \delta$, and $V(t) \le \eta/2$ ensures our temporary assumption, from which we get
$$\left|\frac{\nu_i(t)}{\psi_i(t)}\right| \le 1 - \delta, \quad \forall i \in \cN,$$
hence we have~\eqref{eq:V} and~\eqref{eq:U}, when we are inside the corresponding set.

So, assume that $U(t) = \delta_\eta/\sqrt{\lambda_N}$ and $V(t) \le \eta/2$, then we get from~\eqref{eq:U},
\begin{align*}
\dot{U} &\le - \frac{\underline{\gamma}\lambda_2}{2\eps}\frac{\delta_\eta}{\sqrt{\lambda_N}} + \frac{M_U\sqrt{\lambda_N}}{2} < 0.
\end{align*}
On the other hand, if $U(t) \le \delta_\eta/\sqrt{\lambda_N}$ and $V(t) = \eta/2$, then we have from~\eqref{eq:V},
\begin{align*}
\dot{V} &\le -c\frac{\eta}{2} + M_V\eps + L_\mu \sqrt{\lambda_N}\frac{\delta_\eta}{\sqrt{\lambda_N}} < 0,
\end{align*}
which makes the set positively invariant.

\subsection{Asymptotic convergence}\label{app:asymp}

Now, if we have~\eqref{eq:V} and~\eqref{eq:U} for all $t \ge t_0$ with the performance functions $\psi_i$ such that $\lim_{t \to \infty} \psi_i(t) = 0$, $i \in \cN$, then we can show that $\lim_{t\to \infty} V(t) = 0$, hence
$$\lim_{t \to \infty} |x_i(t) - \xi(t)| = 0, \quad \forall i \in \cN.$$

In particular, we first have $\lim_{t \to \infty} U(t) = 0$ because otherwise, there exists $\underline{U} > 0$ such that $U(t) \ge \underline{U}$ for all $t \ge t_0$, which is a contradiction since we have
\begin{align*}
\dot{U} &\le \frac{M_U\sqrt{\lambda_N}}{2} - \left(\frac{\underline{\gamma}\lambda_2}{\widehat{\psi}(t)} - \lambda_\psi\right)\underline{U} < -1
\end{align*}
whenever $U \ge \underline{U}$ for all $t \ge T$ with some finite but sufficiently large $T \ge t_0$ because $\lim_{t \to \infty} 1/\widehat{\psi}(t) = \infty$.

Then, similarly, we can conclude that $\lim_{t\to \infty}V(t) = 0$ because otherwise, there exists $\underline{V} > 0$ such that $V(t) \ge \underline{V}$ for all $t \ge t_0$, which is a contradiction since we have
\begin{align*}
\dot{V} &\le -c\underline{V} + M_V\widehat{\psi}(t) + L_\mu\sqrt{\lambda_N}U(t) < -1
\end{align*}
whenever $V \ge \underline{V}$ for all $t \ge T$ with some finite but sufficiently large $T \ge t_0$ because $\lim_{t\to \infty} \widehat{\psi}(t) = 0$ and $\lim_{t \to \infty} U(t) = 0$.

\subsection{Proof of Lemma \ref{lem:vu}}\label{app:Lemma}

From \eqref{eq:omega_id} and \eqref{eq:nupsi}, it is seen that 
\begin{align*}
    &\left|\mu_i\left(\frac{\underline{\psi}}{\psi_i} R_i y + \mu_i^{-1}(f_{\text{em}}^s - f_i^s)\right) - \mu_i\left(\mu_i^{-1}(f_{\text{em}}^s - f_i^s)\right)\right| \\
    &\le L_\mu \left|\frac{\underline{\psi}}{\psi_i} R_i y\right| \le L_\mu \|y\|.
\end{align*}
Then, we have by \eqref{eq:y} and \eqref{eq:deriv_x_s},
\begin{align}\label{eq:ineq_x_s}
\begin{split}
|\dot{x}_s| &\le M_f + \frac{L_f}{N} \sum_{i=1}^N |x_i - x_s| + L_\mu \|y\| \\
&\le M_f + \frac{L_f \sqrt{N}}{\lambda_2} \overline{\psi} + 2 r_\psi \sqrt{N} L_\mu =: M_s .
\end{split}
\end{align}
Similarly, 
\begin{align*}
\dot{V} &= \frac{x_s - \xi}{|x_s - \xi|}(\dot{x}_s - \dot{\xi}) \le -c V + M_V \widehat{\psi}(t) + L_\mu \|y\|.
\end{align*}
which comes from the fact that the emergent dynamics~\eqref{eq:emer} is contractive so that $(x_s - \xi)(f_{\text{em}}(t, x_s) - f_{\text{em}}(t, \xi)) \le - c |x_s - \xi|^2$.
This proves \eqref{eq:V}.

Now, let $W := U^2 = y^\top\Lambda^{-1}y$.
Then,
\begin{align*}
&\dot{W} \le 2\left|\frac{\dot{\underline{\psi}}}{\underline{\psi}}\right|\!\left[W \!+\! \frac{1}{\underline{\psi}} \left|y^\top\Lambda^{-1}R^\top\begin{bmatrix} 
\psi_1\mu_1^{-1}(f_{\text{em}}^s - f_1^s) \\ \vdots \\ \psi_N\mu_N^{-1}(f_{\text{em}}^s - f_N^s)\end{bmatrix}\right|\right] \\
& \!+\! \frac{2}{\underline{\psi}} \|y\| \sqrt{\sum_{i=1}^N|f_i(t, x_i) - f_i(t, x_s)|^2} \\
& \!-\! \frac{2}{\underline{\psi}}\sum_{i=1}^N R_iy \! \left[\mu_i\!\begin{pmatrix}\frac{\underline{\psi}}{\psi_i}R_iy \!+\! \mu_i^{-1}\!(f_{\text{em}}^s \!-\! f_i^s)\end{pmatrix} \!-\! \mu_i\!\left(\mu_i^{-1}\!(f_{\text{em}}^s \!-\! f_i^s)\right)\right] \\
& \!+\! 2\|\Lambda^{-1}y\|\sqrt{N}\max_i \!\left|\frac{\psi_i}{\underline{\psi}}\frac{d}{dt}\mu_i^{-1}\!(f_{\text{em}}(t, x_s(t)) \!-\! f_i(t, x_s(t)))\right| \\ 
&\!+\! \frac{2}{\underline{\psi}^2}\|\Lambda^{-1}y\|\sqrt{N}2\lambda_\psi \widehat{\psi}(t)^2 \\
&\le \frac{2N}{\lambda_2}\left(\lambda_\psi + L_f\right) r_\psi \|y\| + 2\lambda_\psi W + \frac{4\sqrt{N}}{\lambda_2}\lambda_\psi r_\psi^2 \|y\| \\
& \!-\! \frac{2}{\underline{\psi}}\sum_{i=1}^N R_iy \!\left[\mu_i\!\begin{pmatrix}\frac{\underline{\psi}}{\psi_i}R_iy \!+\! \mu_i^{-1}\!(f_{\text{em}}^s \!-\! f_i^s)\end{pmatrix} \!-\! \mu_i\!\left(\mu_i^{-1}\!(f_{\text{em}}^s \!-\! f_i^s)\right)\right] \\
& \!+\! 2\|y\|\frac{\sqrt{N}}{\lambda_2}r_\psi\max_i\! \left|\frac{d}{dt}\mu_i^{-1}\!(f_{\text{em}}(t, x_s(t)) \!-\! f_i(t, x_s(t)))\right| 
\end{align*}
where the first inequality follows from the identity:
\begin{align*}
&-y^\top R^\top\boldsymbol{x} = y^\top\Lambda^{-1}\left(-\Lambda R^\top\boldsymbol{x}\right) = \underline{\psi}(t)y^\top\Lambda^{-1}y \\
& + y^\top\Lambda^{-1}R^\top{\rm col}(\psi_1\mu_1^{-1}(f_{\text{em}}^s - f_1^s), \dots, \psi_N\mu_N^{-1}(f_{\text{em}}^s - f_N^s)).
\end{align*}

Now, note that, for $a \not = 0$,
\begin{align*}
&(\mu_i^{-1})'(a) = \frac{1}{\mu_i'(\mu_i^{-1}(a))}  \\
&= \frac{1}{\gamma_i'(|\mu_i^{-1}(a)|)|\mu_i^{-1}(a)| + \gamma_i(|\mu_i^{-1}(a)|)} \le \frac{1}{\gamma_i(0)} \le \frac{1}{\underline{\gamma}}
\end{align*}
where the first equality follows by differentiating $\mu_i(\mu_i^{-1}(a)) = a$ and $\underline{\gamma} = \min_i\gamma_i(0)$.
The assumption that $\gamma_i$ is non-decreasing is utilized for this derivation.
Then, by the above inequality, we now have
\begin{align*}
&\left|\frac{d}{dt}\mu_i^{-1}(f_{\text{em}}(t, x_s(t)) - f_i(t, x_s(t)))\right| \\
&= (\mu_i^{-1})'(f_{\text{em}}^s- f_i^s) \cdot \left|\frac{d}{dt}\left[f_{\text{em}}(t, x_s(t)) - f_i(t, x_s(t))\right]\right| \\
&\le \frac{1}{\underline{\gamma}} \cdot \left[2 L_f |\dot{x}_s(t)| + 2 \theta_f + \frac{\lambda_\psi}{\overline{\gamma}}\right]
\end{align*}
where $\overline{\gamma} := \min_{|a| \le 1 - 2\delta}\min_j(\mu_j^{-1})'(a) > 0$, and \eqref{eq:chi} may be helpful to digest the inequality.

Now, note that $\mu_i(b) - \mu_i(a) = \mu_i'(c)(b - a)$ with some $c \in (a, b)$ by the mean value theorem.
Since $\mu_i'(c) = \gamma_i'(|c|)|c| + \gamma_i(|c|) \ge \gamma_i(0) \ge \underline{\gamma}$, we have 
\begin{align*}
(b - a)(\mu_i(b) - \mu_i(a)) \ge \underline{\gamma}(b - a)^2
\end{align*}
for all $-\infty < a \le b < \infty$.
Therefore, we finally obtain
\begin{align}\label{eq:mu2}
&\dot{W} \le \frac{2N}{\lambda_2}\left(\lambda_\psi + L_f\right) r_\psi \|y\| + 2 \lambda_\psi W - \frac{2}{\widehat{\psi}}\sum_{i=1}^N\underline{\gamma}(R_iy)^2  \nonumber \\
& + 2\|y\| \frac{\sqrt{N}}{\lambda_2} \frac{r_\psi}{\underline{\gamma}} \left[2 L_f M_s + 2\theta_f + \frac{\lambda_\psi }{\overline{\gamma}} + 2\lambda_\psi r_\psi \underline{\gamma} \right] \nonumber \\
&=: M_U \|y\| + 2\lambda_\psi W - \frac{2}{\widehat{\psi}} \underline{\gamma} \|y\|^2 \\
&\le M_U \sqrt{\lambda_N W} + 2\lambda_\psi W - \frac{2}{\widehat{\psi}} \underline{\gamma}\lambda_2 W \nonumber 
\end{align}
where we used $\sum_{i=1}^N (R_i y)^2 = y^\top R^\top Ry = y^\top y$.
This proves~\eqref{eq:U}.
\QEDB

\section{Proof of Lemma~\ref{lem:med}}\label{ssec:proof_lemma_median}

We will prove that if $\left|h\right|_{\mathcal{M}_{\mathcal{F}}} > \eta$, then we have a contradiction.
So, without loss of generality assume that
$$h > f + \eta, \quad \forall f \in \mathcal{M}_{\mathcal{F}}.$$
This ensures that the index set $\mathcal{K} \subset\mathcal{N}$, which consists of all the indexes $i \in \mathcal{N}$ such that $h - f_i>  \eta$, satisfies
$$\sum_{i \in \mathcal{K}} \psi_i > \frac{1}{2}\sum_{i\in \mathcal{N}} \psi_i,$$
according to the definition of $\mathcal{M}_{\mathcal{F}}$.
Now, by the constraint~\eqref{eq:beta_bound}, we have
$$\mu_i^{-1}(h - f_i) \ge 1 - \eps, \quad \forall i \in \mathcal{K},$$
and this gives
\begin{align*}
0 &= \sum_{i=1}^N \psi_i\mu_i^{-1}(h - f_i) \ge \sum_{i \in \mathcal{K}} \psi_i (1 - \eps)  - \sum_{i \in \mathcal{N}\setminus \mathcal{K}}  \psi_i \\
&=  ( 2 - \eps)\sum_{i \in \mathcal{K}} \psi_i - \sum_{i \in \cN} \psi_i \\
&\ge \left[(2 - \eps)\left(\frac{1}{2} + \delta\right) - 1\right]\sum_{i \in \cN} \psi_i > 0
\end{align*}
where the last term is positive by the definition of $\eps$.

\section{Reasoning of the conjecture}\label{app:conj}

In this section, we show that for particular cases, the conjecture can be proved.
For instance, we consider the case when $N$ is odd, $\psi_i = \psi$ for all $i \in \cN$, and the graph is complete and unitary.
For this case, $\mathcal{M}_\chi$ is just a singleton that consists of the median, and thus, by letting $V(t) := \text{med}_{i} x_i(t)$, we have
\begin{align*}
\left|\dot{V}(t)\right| &\le |f_{J(t)}(t, V(t))| + \left| \mu_{J(t)}\left(\frac{\nu_{J(t)}(t)}{\psi(t)}\right) \right|,
\end{align*}
where $J(t) \in \cN$ is such that $x_{J(t)}(t) = V(t)$ and $|\dot{x}_{J(t)}(t)| \ge |\dot{x}_j(t)|$ for any $j \in \cN$ satisfying $x_j(t)  =V(t)$.
The second term is bounded by a constant, which is independent of the function $\psi$ since $\nu_i = \sum_{j=1}^N x_j - Nx_i$, which implies $\nu_{J(t)}(t) = \text{med}_i \nu_i(t)$, and thus, we have
$$ \left|\frac{\nu_{J(t)}(t)}{\psi(t)} \right| \le  \frac{N-1}{N+1} < 1,$$
by the algebraic constraint $\sum_{i = 1}^N \nu_i(t) \equiv 0$.
This is because, if say $\nu_{J(t)}(t)/\psi(t) > (N-1)/(N+1)$, then
$$0 \equiv \sum_{i=1}^N \frac{\nu_i(t)}{\psi(t)} > \frac{N+1}{2}\frac{N-1}{N+1} - \frac{N-1}{2} = 0,$$
which is a contradiction.
Therefore, regardless of the choice of the performance function $\psi$, we have
$$\lim_{t \to t_0} |V(t)|_{\mathcal{M}_\chi} = 0,$$
and thus, by noting that $x_{J(t_1 + \tau)}(t_1+\tau)$ approximates $(1/N)\sum_{i=1}^N x_i(t_1 + \tau)$ for sufficiently small $\eps$, we can conclude that 
$$\lim_{t_1 + \tau \to t_0} \lim_{\eps \to 0} \left|\frac{1}{N}\sum_{i=1}^N x_i(t_1 + \tau) \right|_{\mathcal{M}_\chi} = 0.$$

Another case is when $(1-\eta)\psi_{i^*} > \sum_{j \neq i^*} \psi_j$ with some $\eta > 0$ as in Section~\ref{subsec:elect}.
As shown in Section~\ref{subsec:elect}, we have $|\mu_{i^*}(\nu_{i^*}(t)/\psi_{i^*}(t))| \le M_\eta$ regardless of the choice of the performance functions $\psi_i$, and therefore, can again conclude as
$$\lim_{t_1+\tau \to t_0} \lim_{\eps \to0} \frac{1}{N}\sum_{j=1}^N x_j(t_1 + \tau) = \lim_{t_1+\tau \to t_0} x_{i^*}(t_1+\tau),$$
where $\mathcal{M}_\chi = \{x_{i^*}(t_0)\}$ by its definition.

\end{document}